\documentclass[useAMS,usenatbib]{mnras}
\usepackage{graphicx}
\usepackage{epstopdf,color,verbatim}
\usepackage{threeparttable}
\usepackage[]{txfonts}

\def\simless{\mathbin{\lower 3pt\hbox
{$\rlap{\raise 5pt\hbox{$\char'074$}}\mathchar"7218$}}}   %< or of order
\def\simmore{\mathbin{\lower 3pt\hbox
{$\rlap{\raise 5pt\hbox{$\char'076$}}\mathchar"7218$}}}   %> or of order
\newcommand{\be}{\begin{equation}}
\newcommand{\ee}{\end{equation}}
\newcommand{\bi}{\begin{itemize}}
\newcommand{\ei}{\end{itemize}}

\newcommand{\fig}[1]{Fig.~\ref{fig:#1}}
\newcommand{\fign}[1]{\ref{fig:#1}}
\newcommand{\sect}[1]{Section~\ref{sect:#1}}
\newcommand{\app}[1]{Appendix~\ref{app:#1}}
\newcommand{\eq}[1]{Eq.~(\ref{eq:#1})}
\newcommand{\comp}{c/\omega_{\rm p}}
\newcommand{\omp}{\omega_{\rm p}}
\newcommand{\ompc}{\,\omega_{\rm p}/c}
\newcommand{\xsh}{x_{\rm sh}}
\newcommand{\delgam}{\Delta\gamma}
\newcommand{\ex}[1]{10^{-#1}}

\usepackage{epsfig} 
\usepackage{epstopdf}

\title[]
{The synchrotron maser emission from relativistic magnetized shocks: Dependence on the pre-shock temperature}

\author[A-N.Babul \& L. Sironi]
{Aliya-Nur Babul$^{1}$\thanks{E-mail: aliya.babul@columbia.edu}\ and Lorenzo Sironi$^1$\thanks{E-mail: lsironi@astro.columbia.edu} \\
$^1$Department of Astronomy, Columbia University, 550 W 120th St, New York, NY 10027, USA}

\begin{document}
\date{Received / Accepted}
\pagerange{\pageref{firstpage}--\pageref{lastpage}} \pubyear{2018}

\maketitle

\label{firstpage}

\begin{abstract}
Electromagnetic precursor waves generated by the synchrotron maser instability at relativistic magnetized shocks have been recently invoked to explain the coherent radio emission of Fast Radio Bursts. By means of two-dimensional particle-in-cell simulations, we explore the properties of the precursor waves in relativistic electron-positron perpendicular shocks as a function of the pre-shock magnetization $\sigma\gtrsim 1$ (i.e., the ratio of incoming Poynting flux to particle energy flux) and thermal spread $\delgam\equiv kT/mc^2=\ex{5}-\ex{1}$. We measure the fraction $f_\xi$ of total incoming energy that is converted into precursor waves, as computed in the post-shock frame. At fixed magnetization, we find that $f_{\xi}$ is nearly independent of temperature as long as $\delgam\lesssim \ex{1.5}$ (with only a modest decrease of a factor of three from $\delgam=\ex{5}$ to $\delgam=\ex{1.5}$), but it drops by nearly two orders of magnitude for $\delgam\gtrsim\ex{1}$. At fixed temperature, the scaling with magnetization $f_\xi\sim 10^{-3}\,\sigma^{-1}$ is consistent with our earlier one-dimensional results. For our reference $\sigma=1$, the power spectrum of precursor waves is relatively broad (fractional width $\sim 1-3$) for cold temperatures, whereas it shows pronounced line-like features with fractional width $\sim 0.2$ for $\ex{3}\lesssim\delgam\lesssim \ex{1.5}$. For $\sigma\gtrsim 1$, the precursor waves are beamed within an angle $\simeq \sigma^{-1/2}$ from the shock normal (as measured in the post-shock frame), as required so they can outrun the shock. Our results can provide physically-grounded inputs for FRB emission models based on maser emission from relativistic shocks.
%We calculate the Poynting flux and spectra of the electromagnetic wave produced by magnetized relativistic shocks. Magnetar flare models suggest that the electromagnetic wave produced at the shock front could account for the Fast Radio Burst (FRB) emission that has been observed. In this study we employ Particle-In-Cell simulations in 2D geometry to study the evolution of the shock. We consider a large set of upstream temperatures that we explore ($kT/mc^2$). At the shock front particles form a ring distribution in momentum space which is unstable to the synchrotron maser instability and its collapse results in the emission of an electromagnetic precursor wave. We show that the Poynting efficiency of the steady-state precursor wave is higher in a cold plasma ($kT/mc^2 = 3 \times 10^{-5}$) by a factor of three as compared to the hottest plasma ($kT/mc^2 = 1 \times 10^{-2}$. We further demonstrate that the power spectra of the electromagnetic precursor displays broadband features in cold plasmas as compared to hot plasmas which present sharply peaked spectra. The significant dependence of spectral shape on temperature has the potential to help answer questions about the variety of spectra shapes that have been seen in observed FRBs.
\end{abstract} 


\section{Introduction}
Relativistic shocks are invoked as candidate sources of non-thermal particles in pulsar wind nebulae, gamma-ray bursts, and active galactic nuclei jets, and as possible accelerators of ultra-high-energy cosmic rays  \citep[e.g.][]{blandford_eichler_87}. However, relativistic shocks are generally quasi-perpendicular, i.e., with pre-shock field orthogonal to the shock direction of propagation, a configuration that --- if the magnetic field is sufficiently strong --- inhibits efficient particle acceleration \citep[e.g.][]{begelman_kirk_90,sironi_13}. While poor particle accelerators, relativistic magnetized perpendicular shocks can channel an appreciable fraction of the incoming flow energy into semi-coherent electromagnetic waves propagating back into the upstream medium (hereafter, ``precursor waves'' moving ahead of the shock). The waves are attributed to the synchrotron maser instability \citep{alsop_arons_88,hoshino_91}. The instability is sourced by a population inversion that naturally occurs at the shock front, where a coherent ring-like distribution is constantly produced as the incoming particles gyrate in the shock-compressed field. 

 Recently, the discovery of Fast Radio Bursts %\citep[FRBs; ][]{lorimer_07, keane_12, thornton_13, spitler_14,Ravi+15,Champion+16,Shannon+18} 
 \citep[FRBs; for recent reviews, see][]{petroff_19,cordes_19,platts_19}
 has revived the interest in the synchrotron maser. FRBs are bright ($\sim 1\,$Jy) pulses of millisecond duration detected in the $\sim$~GHz band. Their extremely high brightness temperature, $T_{\rm B} \sim 10^{37}\,$K, requires a coherent emission mechanism \citep[e.g.,][]{katz_16}. The synchrotron maser at relativistic shocks has been invoked as one of the candidate emission mechanisms within the so-called ``magnetar scenario'' \citep{lyubarsky_14,murase_16,belo_17,waxman_17,lorenzometzger,belo_19,margalit_20,margalit_20b}, where magnetars are invoked as FRB progenitors, a hypothesis recently confirmed by the detection of FRBs from a Galactic magnetar \citep{scholz_20,bochenek_20}. In response to motions of the magnetar crust, the above-lying magnetosphere is violently twisted and a strongly magnetized pulse is formed, which propagates away through the  magnetar wind. In the shock maser scenario, the FRB is generated at ultra-relativistic shocks resulting from the collision of the magnetized  pulse with the steady wind  that is produced by the magnetar spin-down luminosity or by the cumulative effect of earlier flares.
 
 The fundamental properties of the precursor waves generated by the synchrotron maser --- i.e., their efficiency, power spectrum, angular distribution and polarization -- can be quantified with self-consistent particle-in-cell (PIC) simulations.  PIC simulations of relativistic magnetized shocks focusing on the synchrotron maser emission have been performed both for electron-positron plasmas \citep[][]{langdon_88,gallant_92, sironi_spitkovsky_09, iwamoto_17, iwamoto_18,plotnikov_18,plotnikov2019} and electron-proton or electron-positron-proton plasmas \citep{hoshino_92,amato_arons_06,lyubarsky_06,hoshino_08,stockem_12,iwamoto_19}. 
 
 These works primarily focused on low magnetizations $\sigma\lesssim 1$, where $\sigma$ is the ratio of upstream Poynting flux to kinetic energy flux. On the other hand, FRBs are expected to originate from extreme environments where the energy content of the plasma is dominated by magnetic fields, as in magnetar winds. In \citet{plotnikov2019}, we performed one-dimensional (1D) PIC simulations of electron-positron shocks and investigated how the properties of the synchrotron maser depend on the flow magnetization, in the $\sigma\gtrsim1$ regime most relevant for FRB sources.  We found that the shock converts a fraction  $ f_\xi \approx 2 \times 10^{-3}\sigma^{-1}$ of the total incoming energy into the precursor waves, as measured in the post-shock (downstream) frame. 
 %In the shock rest frame, the efficiency is $f_\xi' \approx 7 \times 10^{-4}/\sigma^2$. 
 %The peak frequency of the wave power spectrum scales in the post-shock frame as $\omega_{\rm peak} \simeq 3\, \omega_{\rm p}\max[1,\sqrt{\sigma}]$, where $\omega_{\rm p}$ is the plasma frequency. 
 At $\sigma\gtrsim1$, %where our estimated $\omega_{\rm peak}$ differs from earlier works (that quoted $\omega_{\rm peak}\propto \sigma \omega_{\rm p}$, see  \citet{gallant_92},  rather than $\omega_{\rm peak}\propto \sqrt{\sigma} \omega_{\rm p}$ as we find)
 we showed that the shock structure displays two solitons separated by a cavity, and the peak frequency of the spectrum corresponds to an eigenmode of the cavity. We also found that the efficiency and spectrum of the precursor waves do not depend on the bulk Lorentz factor of the pre-shock flow.
 
 The results in \citet{plotnikov2019} were obtained assuming that the pre-shock flow has small thermal spread, $\delgam\equiv kT/m c^2=\ex{4}$. In fact, with the exception of the study by \citet{amato_arons_06} --- who focused on nonthermal lepton acceleration in pair-proton plasmas, rather than on the properties of the precursor waves --- all prior studies were conducted in the limit of negligible upstream temperatures. In this work, we discuss the dependence of the precursor waves generated by the synchrotron maser on the upstream temperature, by means of two-dimensional (2D) PIC simulations of relativistic magnetized electron-positron shocks. We focus on magnetically-dominated plasmas ($\sigma=1$ and 3) and explore thermal spreads in the range $\delgam=\ex{5}-\ex{1}$. All our simulations are evolved for sufficiently long ($\gtrsim 4000\,\omp^{-1}$) so that the properties of the precursor waves, such as their Poynting flux and power spectrum, attain a steady state. At fixed magnetization, we find that the efficiency $f_{\xi}$ is nearly independent of temperature as long as $\delgam\lesssim \ex{1.5}$ (with only a modest decrease of a factor of three from $\delgam=\ex{5}$ to $\delgam=\ex{1.5}$), but it drops by nearly two orders of magnitude for $\delgam\gtrsim\ex{1}$. For our reference $\sigma=1$, the power spectrum of precursor waves is relatively broad (fractional width $\sim 1-3$) for cold temperatures, whereas it shows narrow line-like features with fractional width $\sim 0.2$ for $\ex{3}\lesssim\delgam\lesssim \ex{1.5}$. For $\sigma\gtrsim 1$, the precursor waves are beamed within an angle $\simeq \sigma^{-1/2}$ from the shock normal (as measured in the post-shock frame), as required so they can outrun the shock. %Within the shock maser scenario for FRBs, our results can help constraining the properties of magnetar winds. 
 
The paper is organized as follows. In Section~\ref{sect:setup} we present the numerical method and the simulation setup. We then discuss the main results of our investigation, as regard to shock structure (\sect{shock}), precursor efficiency (\sect{eff}) and beaming and power spectrum (\sect{spec}). We summarize our findings in \sect{disc} and discuss their astrophysical implications.
 
  %Considerable work has been done to study the dependence of the precursor emission on the background plasma magnetization. In both 1D and 2D simulations of magnetizations ranging from $1 \times  10^{-3}$ to $3  \times  10^{-1}$ with a magnetic field perpendicular to the simulation plane the precursor wave energy was demonstrated to be a few percent of the upstream magnetic field energy and was shown to decrease with increasing magnetization\citep{hoshino_92,gallant_92,hoshino_08,iwamoto_17}. Simulations of perpendicular shocks with magnetic fields in the simulation plane show the generation of X-modes wave as the result of the SMI, as well as O-mode waves. For magnetizations ranging from $3 \times  10^{-3}$ to 1 both the X-mode and O-mode precursor wave magnetic field, normalized to the upstream magnetic field, was shown to decrease with increasing magnetization. For small magnetizations ($\sigma < 1 \times 10^{-2}$) the O-mode wave energy is larger than the the X-mode wave energy while the converse is true for large magnetizations ($\sigma > 1 \times 10^{-2}$) \citep{iwamoto_18}. Studies examining the power spectra of the precursor emission have shown the peak frequency of the precursor emission to roughly follow an increasing trend with increasing values of magnetization \citep{gallant_92,iwamoto_17,iwamoto_18}. 

%%%%%%%%%%%%%%%%%%%%%
%%%%%%%%%%%%%%%%%%%%%
%%%%%%%%%%%%%%%%%%%%%
\section{Simulation Setup}
\label{sect:setup}

We use the three-dimensional (3D) electromagnetic PIC code TRISTAN-MP \citep{spitkovsky_05} to perform simulations of relativistic magnetized shocks in pair plasmas. We perform simulations in 2D spatial domains, but all three components of particle velocities, electric currents and electromagnetic fields are retained.

The simulations are performed in the post-shock frame. The upstream flow, consisting of electrons and positrons, drifts in the $-\hat{x}$ direction with speed $-\beta_0\hat{x}$, where $\beta_0=(1-1/\gamma_0^2)^{1/2}$. For the simulations presented in this paper, we employ $\gamma_0=10$, but we have verified that larger values of $\gamma_0$ (up to $\gamma_0=80$) do not change our conclusions (see also \citet{plotnikov2019}, for an investigation of the dependence on $\gamma_0$ with 1D simulations). The incoming flow reflects off a wall at $x=0$. The shock is formed by the interaction of the incoming and reflected flows and propagates along $+\hat{x}$. The upstream temperature is cast in terms of the thermal spread $\delgam\equiv kT/mc^2$, which we vary from $\delgam=\ex{5}$ up to $\delgam=\ex{1}$. Here, $m$ is the electron mass and $c$ is the speed of light.

The pre-shock plasma carries a frozen-in magnetic field $\bmath{B}_0=B_0\,\hat{z}$ orthogonal to the $xy$ plane of our simulations, and the associated motional electric field $\bmath{E}_0=-\beta_0 B_0\,\hat{y}\equiv - E_0\,\hat{y}$. Our  field configuration parallels the one employed by \citet{iwamoto_17}, and corresponds to a perpendicular shock with out-of-plane upstream field. The magnetic field strength is parameterized via the magnetization, i.e., the ratio of  upstream Poynting flux to kinetic energy flux,
\be
\sigma = {B_{0}^2 \over 4 \pi \gamma_0 N_0 m c^2} = \left( {\omega_{\rm c} \over  \omega_{\rm p}} \right)^2=\left( {\comp \over  r_{\rm L}} \right)^2 \, ,
\label{eq:define_sigma}
\ee
where $N_0$ is the number density of the upstream plasma (including both species), $\omega_{\rm c}=eB_0/\gamma_0 m c$ is the Larmor frequency, $\omp=(4 \pi N_0 e^2/\gamma_0 m)^{1/2}$ is the plasma frequency, $\comp$ is the plasma skin depth and $r_{\rm L}=\gamma_0 m c^2/e B_0$ is the Larmor radius of particles with Lorentz factor $\gamma_0$. Here, $e$ is the electron charge. We explore two values of magnetization, $\sigma=1$ and 3.
We use the plasma skin depth $\comp$ as our unit of length and the inverse plasma frequency $\omp^{-1}$ as our unit of time.

We employ periodic boundary conditions in the $y$ direction. The incoming plasma is injected through a ``moving injector,'' which moves along $+{\hat{x}}$ at the speed of light. The simulation box is expanded in the $x$ direction as the injector approaches the right boundary of the computational domain. This permits us to save memory and computing time, while following the evolution of all the upstream regions that are causally connected with the shock \citep[for details see, e.g.][]{spitkovsky_05,sironi_spitkovsky_09}. Over time, the distance between the shock and the injector increases, and the incoming flow might suffer from the so-called numerical Cerenkov instability \citep[e.g.][]{dieckmann_06}. By employing a fourth-order spatial stencil for Maxwell's equations \citep{greenwood_04}, we find no evidence of numerical Cerenkov instability within the timespan covered by our simulations.

The leftmost edge of the downstream region, which is a conducting boundary for electromagnetic fields and a reflecting wall for particles (hereafter, the ``wall''), is initially located at $x=0$. The focus of this work is on upstream-propagating waves generated by the shock, rather than on the properties of the shocked plasma. In order to save memory and computing time, we choose to periodically jump the wall toward the shock, such that the average speed of the wall is $\sim 5\%$ lower than the shock speed and the wall always stays safely behind the shock (by at least a few tens of $\comp$). After every jump, we enforce conducting boundary conditions for the electromagnetic fields at the new position of the wall, and we  discard particles to the left of the wall. By performing a few tests without the ``jumping wall,'' i.e., retaining the whole downstream region, we have verified that this strategy does not artificially affect any property of the precursor waves.\footnote{It may be argued that, by removing a significant portion of the downstream region, one may inhibit the downstream plasma from e.g., relaxing to isotropy (which in turn would affect the shock speed). However, our choice of a perpendicular out-of-plane field is by itself suppressing relaxation to full isotropy (the particles only isotropize in the $xy$ plane perpendicular to the field), regardless of whether we employ the jumping wall or not.}

We now describe the numerical parameters used in our work. We employ a high spatial resolution, with $\comp=100$ cells. This ensures that the high frequency / wavenumber part of the power spectrum of precursor waves (i.e., $k\,\comp\gg1$) is properly captured \citep{iwamoto_17}. A few tests with a lower spatial resolution of $\comp=50$ cells have shown good agreement with our production runs, so a resolution of $\comp=50$ cells might also be sufficient. The transverse size of the box is 1440 cells, corresponding to $\sim 14\,\comp$. This is sufficient to capture genuine 2D effects in the properties of the shock and of the precursor waves (e.g., filamentation of the upstream density, see \sect{shock}). Experiments with even larger boxes give essentially the same results. 

The numerical speed of light is $0.45$ cells/timestep. We evolve our simulations for a few thousands of $\omp^{-1}$, which is sufficient to study the steady-state properties of precursor emission. Our longest simulations have been run for $\sim 6000\,\omp^{-1}$, corresponding to more than 1.3 million timesteps.

Our 2D simulations are typically initialized with $N_0=16$ particles per cell (including both species) for $\sigma=1$ and $4$ particles per cell for $\sigma=3$. For $\sigma=1$, we have also performed simulations with $N_0=4$ for all the temperature values we investigated, finding excellent agreement with our reference $N_0=16$ cases (in fact, in \fig{time} and \fig{xieff} we use $N_0=4$ simulations). The simulations with the hottest upstream plasma, $\delgam=\ex{1}$, employ 32 particles per cell (for both $\sigma=1$ and 3), since the precursor emission is very weak, and so harder to properly capture (see \sect{eff}). For this temperature, we have checked that simulations with even larger $N_0=128$ give the same results. In order to further reduce numerical noise in the simulations, we filter the electric current deposited to the grid by the particles, effectively mimicking the role of a larger number of particles per cell \citep{spitkovsky_05,belyaev_15}. We typically apply $N_{\rm sm}=20$ passes of a binomial low-pass digital filter at every timestep. In \app{num}, we show the effects of $N_0$ and $N_{\rm sm}$ on the precursor wave spectrum.

In addition to 2D simulation, which constitute the bulk of this paper, we have also performed a suite of 1D simulations with $\sigma=1$ and the same range of temperatures as in 2D. In 1D simulations, we typically employ $N_0=40$ particles per cell, a spatial resolution of $\comp=112$ cells and a numerical speed of light of $0.5$ cells/timestep. For the hottest temperature $\delgam=\ex{1}$, the number of particles per cell is increased to $N_0=400$.

%The instability is due to both the grid-Cerenkov instability which arises as a result of the fact that electromagnetic waves on a discrete grid have subluminal phase velocity for large wavenumbers, and spurious plasma oscillations. 

\begin{figure}
\centering
%\begin{minipage}[b]{.45\textwidth}
\includegraphics[width=1.0\columnwidth,angle=0]{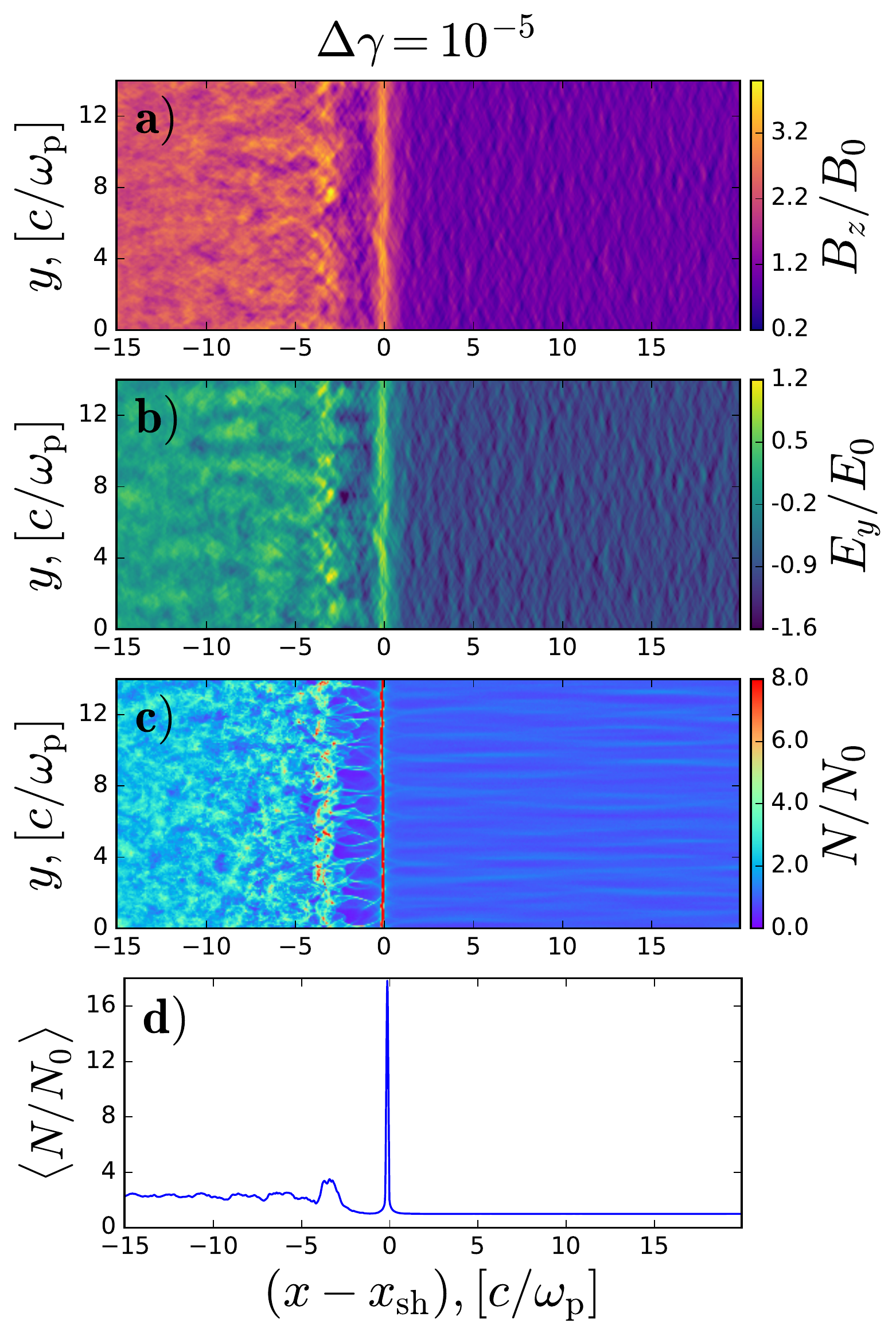}
\vspace{-5mm}
\caption{Shock structure from the 2D PIC simulation with $\sigma=1$ and $\delgam=10^{-5}$ at $\omp t=2000$, when the precursor emission has reached a steady state. We focus on the vicinity of the shock. We present (a) the transverse magnetic field $B_z/B_0$; (b) the transverse electric field $E_y/E_0$; (c) the particle number density, in units of the upstream density $N_0$; (d) the particle number density averaged along the $y$ direction.
}\label{fig:shock5}
    \vspace{-5mm}
\end{figure}

\begin{figure}
    \centering
    \includegraphics[width=\columnwidth]{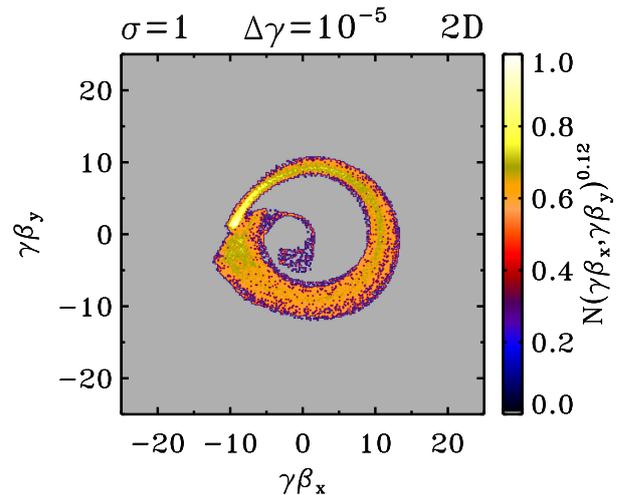}
    \vspace{-5mm}
    \caption{Momentum space $\gamma\beta_x-\gamma\beta_y$ of positrons, from the 2D simulation with $\delgam=\ex{5}$. The particles are selected at time $\omp t=2000$ to be located near the soliton-like structure at the shock front, in the range $-1.5\,\comp<x-x_{\rm sh}<1.5\,\comp$. The histogram is normalized such that $N(\gamma\beta_x,\gamma\beta_y)=1$ in the pixel with the highest value, and the color scale is stretched with $0.12$ power to emphasize weak phase space structures.}
        \vspace{-10mm}
    \label{fig:pspace5}
\end{figure}

\begin{figure}
\centering
%\begin{minipage}[b]{.45\textwidth}
\includegraphics[width=1.0\columnwidth,angle=0]{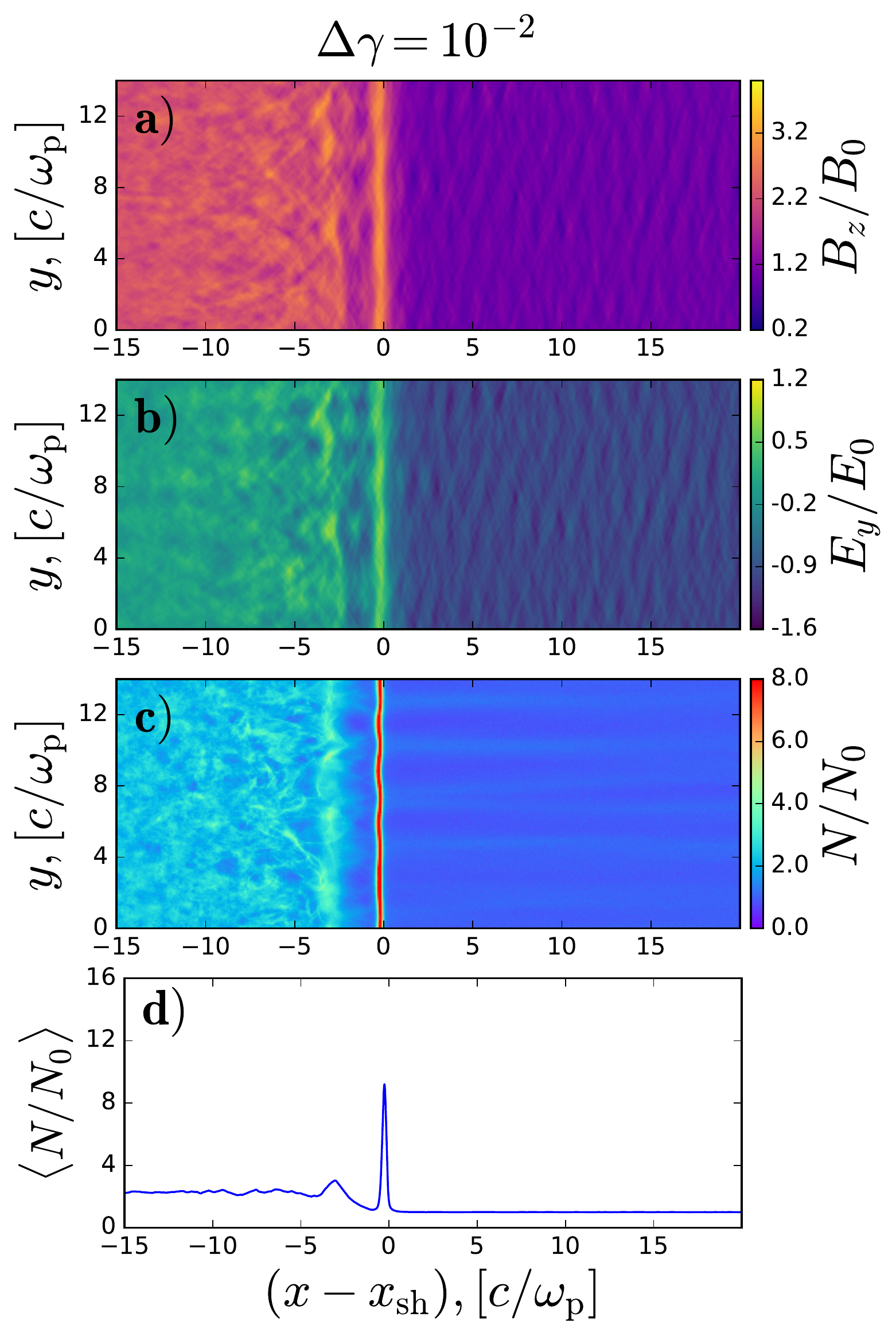}
\caption{Same as in \fig{shock5}, but for a 2D simulation with $\sigma=1$ and upstream plasma temperature of $\delgam=\ex{2}$ at time $\omp t=2000$.}\label{fig:shock2}
%\end{minipage}
\end{figure}

\begin{figure}
    \centering
    \includegraphics[width=\columnwidth]{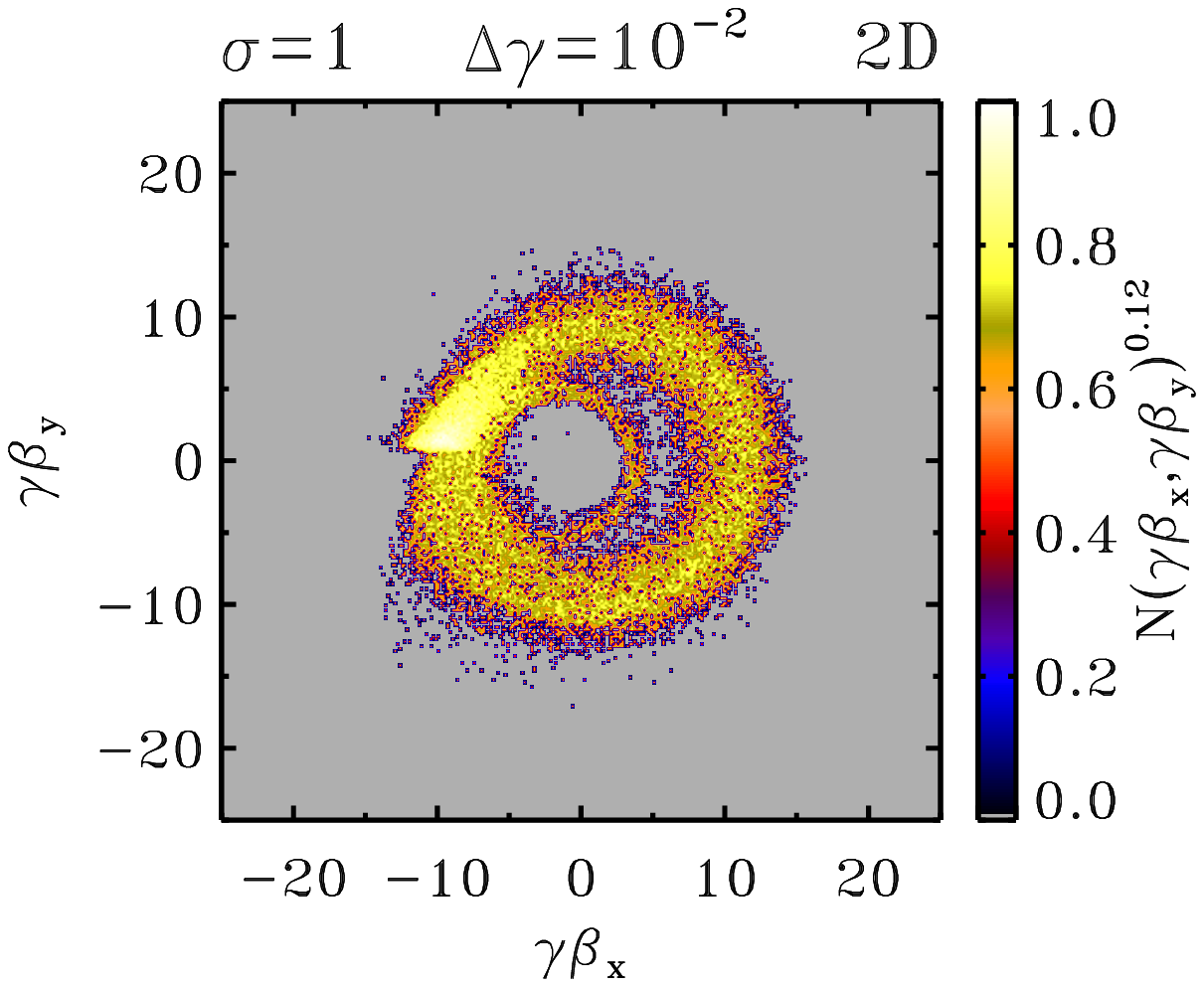}
    \caption{Same as in \fig{pspace5}, but for the 2D simulation with $\delgam=\ex{2}$ at time $\omp t=2000$.}
    \label{fig:pspace2}
\end{figure}

\begin{figure}
\centering
%\begin{minipage}[b]{.45\textwidth}
\includegraphics[width=1.0\columnwidth,angle=0]{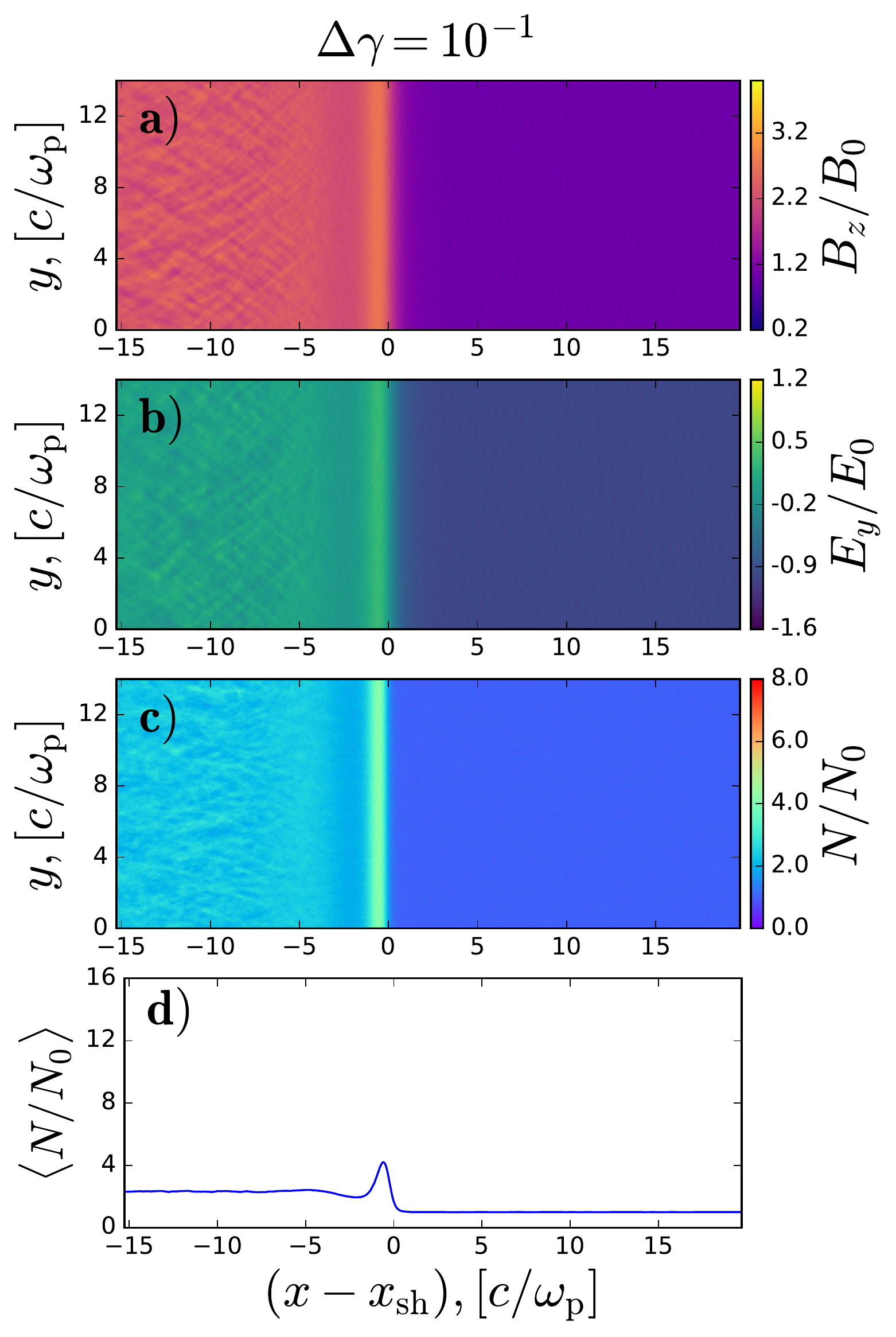}
\caption{Same as in \fig{shock5}, but for a 2D simulation with $\sigma=1$ and upstream plasma temperature of $\delgam=\ex{1}$ at time $\omp t=1500$.}\label{fig:shock1}
%\end{minipage}
\end{figure}

\begin{figure}
    \centering
    \includegraphics[width=\columnwidth]{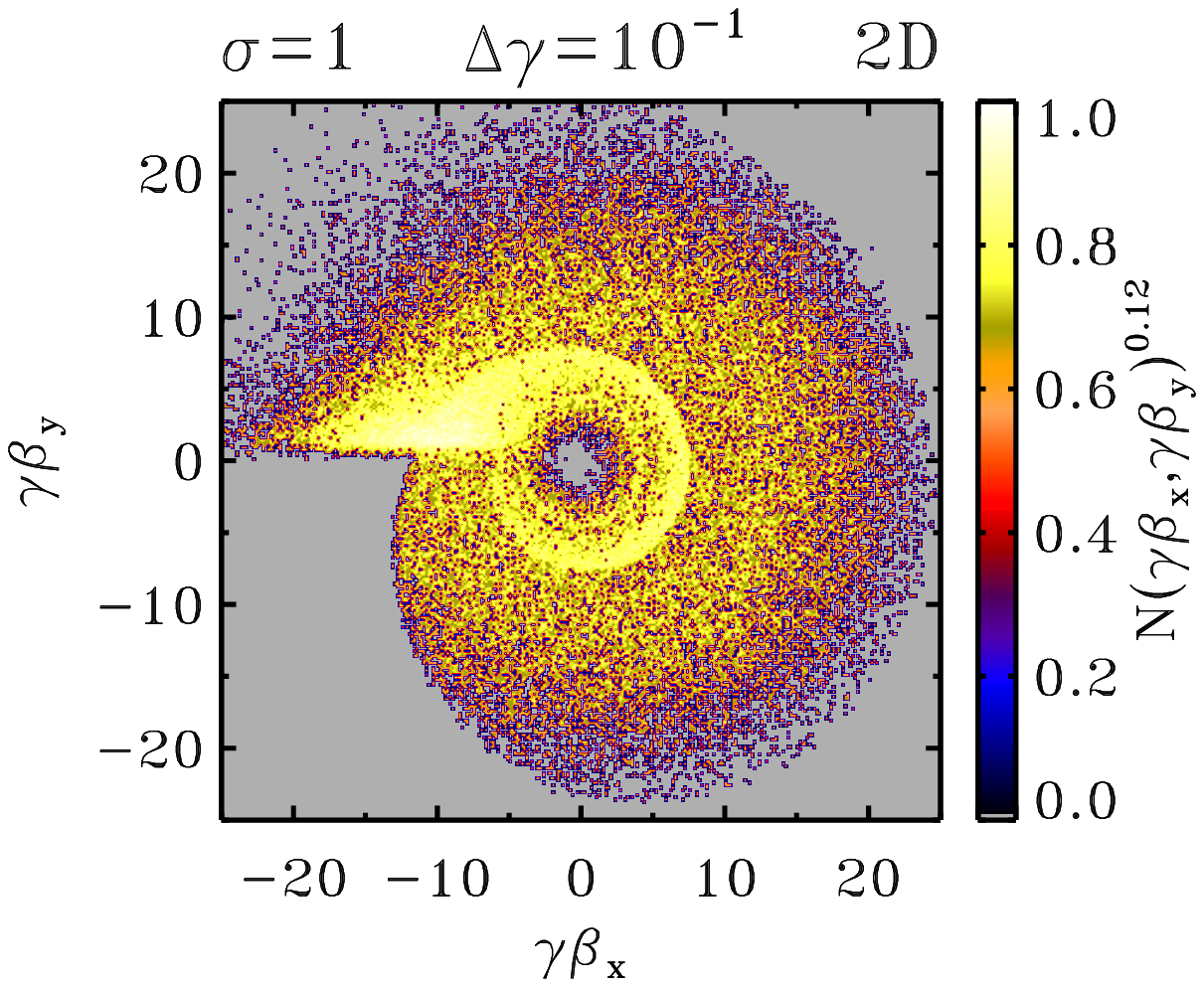}
    \caption{Same as in \fig{pspace5}, but for the 2D simulation with $\delgam=\ex{1}$ at time $\omp t=1500$.}
    \label{fig:pspace1}
\end{figure}

%%%%%%%%%%%%%
%%%%%%%%%%%%%
%%%%%%%%%%%%%
\section{Shock Structure}\label{sect:shock}
In Figs.~\fign{shock5}, \fign{shock2} and \fign{shock1} we present the 2D shock structure from our simulations with $\sigma=1$ and three values of upstream thermal spread: respectively, $\delgam=\ex{5}$, $\ex{2}$ and $\ex{1}$. From top to bottom, in each figure we present the transverse magnetic field $B_z/B_0$, the transverse electric field $E_y/E_0$, the particle number density in units of the upstream density $N_0$,  and the particle number density averaged along the $y$ direction. 

The shock front is at $x=x_{\rm sh}$. The upstream flow is on the positive side $(x-x_{\rm sh}>0)$ and the downstream plasma is on the negative side $(x-x_{\rm sh}<0)$. The existence of a well-developed shock is confirmed by the jump in number density and in $B_z$ at the front location. The shock front itself exhibits a soliton-like structure, as revealed by the density spikes in panels (d) at $x-x_{\rm sh}\sim0$ \citep[see, e.g.,][]{alsop_arons_88}. The density spike in the soliton is higher for colder plasmas (compare panels (d) among the three figures), as derived analytically by \citet{chiueh_91}.
In the soliton, the incoming particles gyrate around the compressed magnetic field and form a semi-coherent ring in momentum space. As shown in Figs.~\fign{pspace5}, \fign{pspace2} and \fign{pspace1}, where we plot, for different values of $\delgam$, the $\gamma\beta_x-\gamma\beta_y$ momentum space of particles  populating the density spike, the thickness of the ring  depends on the pre-shock temperature. A cold well-defined ring appears for low temperatures ($\delgam=\ex{5}$ in \fig{pspace5}), whereas the center of the ring is nearly filled with particles for hot flows ($\delgam=\ex{1}$ in \fig{pspace1}). The radius of the ring is $\sim \gamma_0\beta_0\sim 10$, corresponding to the bulk four-velocity of incoming particles.

The synchrotron maser instability, and the resulting precursor waves, is believed to be sourced by the population inversion in the ring \citep{alsop_arons_88,hoshino_91}.\footnote{We remark that the continuous flow of plasma through the shock ensures that the population inversion is steadily maintained.} Such a population inversion tends to disappear for hot flows, as shown in \fig{pspace1}. We then expect that the synchrotron maser emission will become inefficient in hot plasmas \citep[see also][]{amato_arons_06}. As a result of the synchrotron maser instability, a train of semi-coherent large-amplitude electromagnetic precursor waves is emitted toward the upstream, as shown in the $B_z/B_0$ and $E_y/B_0$ panels, for cold ($\delgam=\ex{5}$ in \fig{shock5}) and moderate ($\delgam=\ex{2}$ in \fig{shock2}) temperatures. As expected, no evidence of precursor waves is seen in hot plasmas ($\delgam=\ex{1}$ in \fig{shock1}). 

When the precursor emission is efficient, electromagnetic waves are seen not only in the upstream region ($x>x_{\rm sh}$), but also right behind the leading soliton, in the density cavity at $-2\,\comp\lesssim x-\xsh\lesssim 0$. As discussed in \citet{plotnikov2019}, this cavity is a peculiarity of $\sigma\gtrsim 1$ shocks. It plays an important role in setting the properties of  precursor waves, since the peak frequency of the wave spectrum is observed to correspond to an eigenmode of the cavity, i.e., the precursor waves might be resonantly amplified by the density cavity. The hot case in \fig{shock1} does not display such a density cavity, and in fact its precursor emission is strongly suppressed (see \sect{eff}). For $\sigma\gtrsim 1$ the precursor waves appear to be generated by an oscillating current localized near the downstream side of the cavity (at $x\sim\xsh -2\,\comp$). This is generally not accounted for within the standard description of the synchrotron maser instability \citep{alsop_arons_88,hoshino_01}.
A characterization of the cavity, and its role in setting the  oscillating current that ultimately drives the precursor waves, is left for future work. 

The wave vector $\bmath{k}$ of the precursor waves is nearly aligned with the shock direction of propagation. The fluctuating magnetic field is along  $z$  (i.e., along the same direction as the upstream field $\bmath{B}_0$), and the fluctuating electric field is perpendicular to both $\bmath{k}$ and $\bmath{B}_0$. The wave is then linearly polarized and identified with the extraordinary mode (X-mode). We remark that 2D simulations with out-of-plane fields do not allow for the excitation of the ordinary mode (O-mode). \citet{iwamoto_18} performed 2D simulations of weakly magnetized shocks ($\sigma\lesssim 1$) with in-plane upstream fields and showed that O modes are stronger than  X modes at low magnetizations ($\sigma\lesssim \ex{2}$), but they become weaker as $\sigma$ increases. 
%In Appendix B of \citet{plotnikov2019}, we discussed the relative importance of O modes and X modes in 3D shock simulations of  cold plasmas ($\delgam=\ex{4}$), and described that O modes are generally suppressed at $\sigma\gtrsim 1$. 
A detailed 3D investigation of the dependence of the precursor emission (in terms of both O and X modes) on the pre-shock temperature will be presented in a forthcoming study.   

Figs.~\fign{shock5} and \fign{shock2} anticipate that the precursor wave spectrum in cold plasmas peaks at higher frequencies / wavenumbers than in warmer plasmas (see \sect{spec} below). This is suggested by comparing panels (a) between \fig{shock5} and \fign{shock2}. Smaller scale structures in $B_z$ are seen for colder plasmas, i.e., the power contained in high-frequency fluctuations is larger in colder plasmas.

Whenever the precursor emission is efficient (i.e., in Figs.~\fign{shock5} and \fign{shock2}, but not \fign{shock1}), filamentary structures are observed in the density (panels (c)), elongated along the shock direction of propagation. Following \citet{iwamoto_17} and \citet{plotnikov_18}, we attribute these structures to the self-focusing and filamentation of the high-amplitude electromagnetic wave when it propagates through the upstream plasma, as studied analytically in electron-proton plasmas by \citet{max_74,drake_74}. The density filaments in the warm case (\fig{shock2}) appear less sharp than in the cold case (\fig{shock5}), suggesting that the precursor waves are not interacting as strongly with the upstream flow. The dominant wavelength is also larger for the warm case than for the cold case, in agreement with the analytical results of \citet{drake_74} (which, however, were obtained for an electron-proton plasma).

\begin{figure}
\includegraphics[width=\columnwidth]{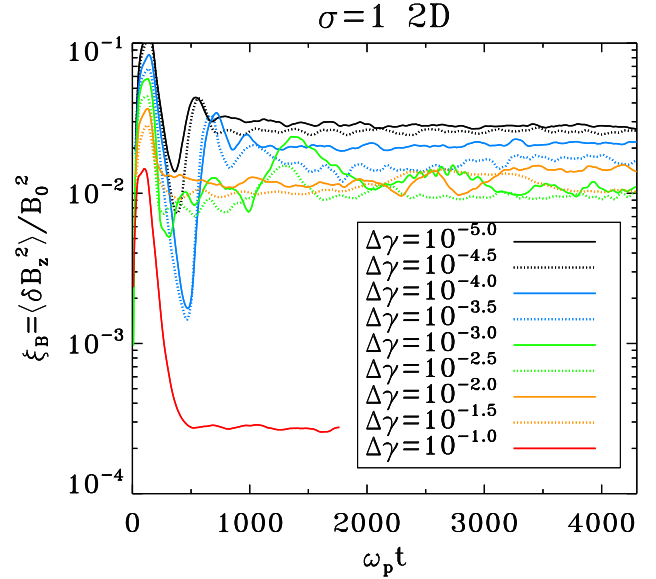}
 \caption{Time evolution of the normalized precursor wave energy, $\xi_B= {\langle \delta B_z^2 \rangle / B_0^2}$, for different values of the upstream thermal spread $\Delta\gamma=kT/m c^2$, as indicated in the legend. The precursor wave energy was extracted from a $25\, c/\omega_{\rm p}$-wide slab located at $5\,c/\omega_{\rm p} < x-x_{\rm sh}<30\, c/\omega_{\rm p}$. The time evolution of the normalized Poynting flux associated to the precursor waves, ${\langle \delta E_y \delta B_z \rangle / E_0B_0}$, is indistinguishable. All the curves in this figure refer to simulations with $N_{0}=4$, apart from the red line ($\delgam=\ex{1}$) which uses $N_0=32$.}
     \label{fig:time}
\end{figure}

\begin{figure}
\centering
    \includegraphics[width=0.95\columnwidth]{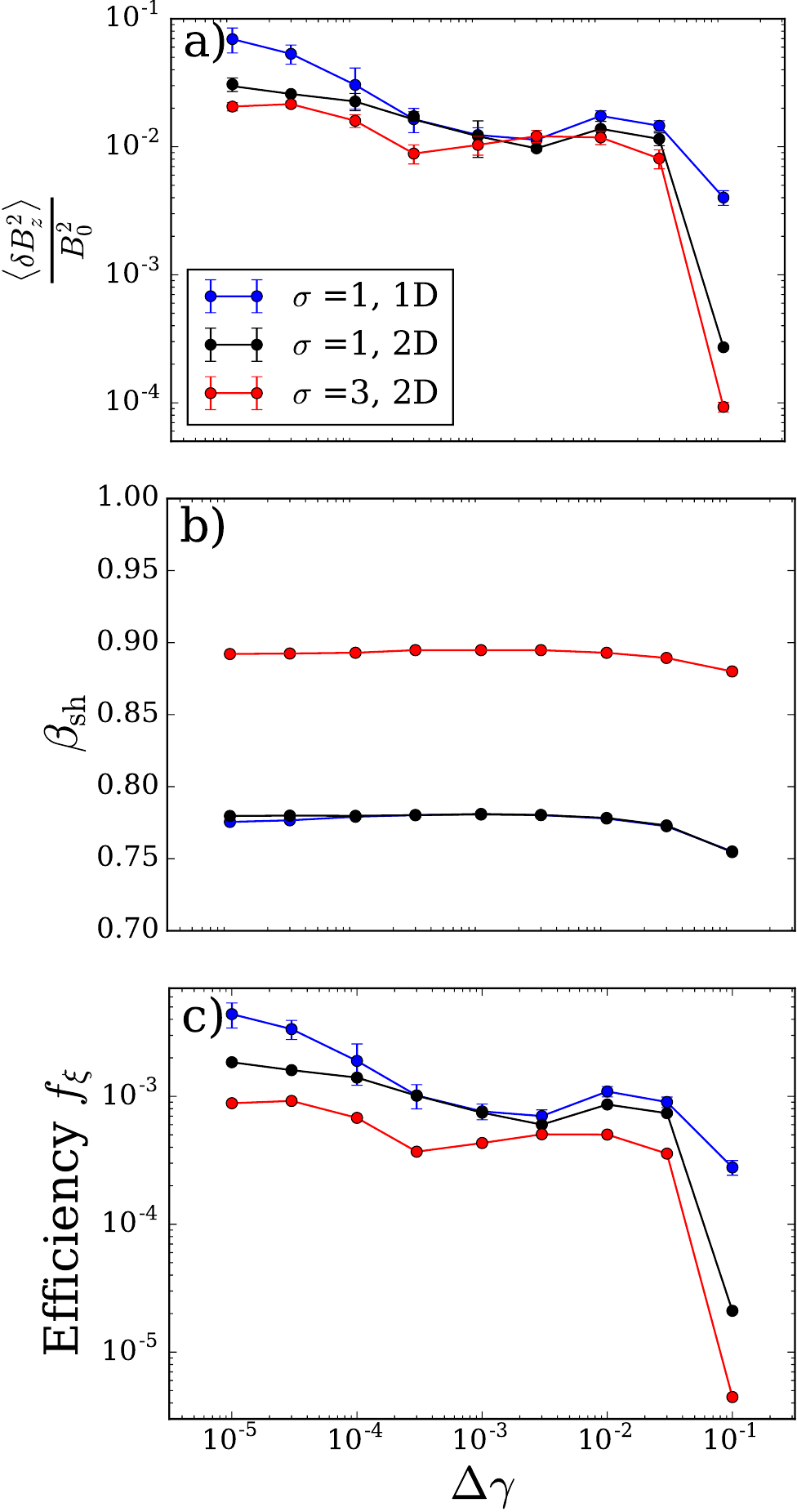}
  \caption{Dependence of the steady-state precursor properties on magnetization, temperature and dimensionality of the simulation domain. In each panel, red points refer to 2D simulations with $\sigma=1$, blue points to 1D simulations with  $\sigma=1$, and red points to 2D simulations with $\sigma=3$ (see legend). 2D simulations employ $N_0=4$, whereas 1D simulations use $N_0=40$, apart from the hottest temperature ($\delgam=\ex{1}$) where 2D runs have $N_0=32$ and 1D runs use $N_0=400$. The data and error bars in each panel are the mean and standard deviation in the time range $\omp t\ge 2000$ for $\sigma=1$ and $\omp t\ge 3000$ for $\sigma=3$. Top panel: normalized precursor wave energy $\xi_B=\langle \delta B_z^2 \rangle /B_0^2$. In all cases, the values of the time-averaged normalized Poynting flux ${\langle \delta E_y \delta B_z \rangle / E_0B_0}$ are indistinguishable. Middle panel: shock velocity in the downstream frame, in units of the speed of light. Bottom panel: precursor efficiency $f_\xi$ measured in the downstream frame, as defined in \eq{xidef}.}
      \label{fig:xieff}
\end{figure}

%%%%%%%%%%%%%%
%%%%%%%%%%%%%%
%%%%%%%%%%%%%%
\section{Precursor Efficiency}\label{sect:eff}
In this Section, we quantify how the wave efficiency depends on the flow temperature and magnetization.
We measure the wave intensity in a region between 5 $c/\omega_{\rm p}$ and 30 $c/\omega_{\rm p}$ ahead of the shock front.\footnote{Our results do not appreciably change if the region is extended further upstream.} This region is far enough from the shock not to be affected by the front structure itself, and it contains a large number of precursor wavelengths so that we can obtain a solid measure of the precursor average properties. The wave intensity is then calculated as the spatial average
\begin{equation}
\langle \delta B_z^2 \rangle =\langle (B_z-B_0)^2 \rangle \, ,
\end{equation}
and we define the normalized wave energy as
\begin{equation}
\xi_B = {\langle \delta B_z^2 \rangle \over B_0^2},
\label{eq:xiB_definition}
\end{equation}
We have verified that in all our simulations $\langle\delta B_z^2 \rangle/B_0^2\simeq\langle\delta E_y\delta B_z \rangle/(E_0B_0)$, i.e., the parameter $\xi_B$ also quantifies the ratio of wave Poynting flux to incoming Poynting flux. Here, $\delta E_y=E_y+\beta_0B_0=E_y+E_0$.

In \fig{time}, we show  for different temperatures the time evolution of the normalized wave energy, for 2D simulations with $\sigma=1$ (see legend). Following a transient, all the curves reach a steady state at $\omp t\gtrsim 1000$. The precursor efficiency is nearly independent of temperature as long as $\delgam\lesssim \ex{1.5}$ (with only a modest decrease of a factor of three from $\delgam=\ex{5}$ to $\delgam=\ex{1.5}$), but between $\delgam=\ex{1.5}$ and $\delgam=\ex{1}$ it drops by nearly two orders of magnitude (red curve). 
We have confirmed this result with 3D simulations (Sironi et al, in prep.).

The steady-state values of the normalized wave energy are shown as a function of temperature in the top panel of \fig{xieff}. There, we present results from 1D and 2D simulations with $\sigma=1$ (blue and black points, respectively) and from 2D simulations with $\sigma=3$ (red points). All the values are extracted from a time range when the precursor has achieved a steady state, more specifically at $\omp t\gtrsim 2000$ for $\sigma=1$ and $\omp t\gtrsim 3000$ for $\sigma=3$. As shown in the top panel of \fig{xieff}, 1D results for $\sigma=1$ generally follow the same trend as our reference 2D simulations (compare blue and black lines), with two notable differences. At very high temperatures ($\delgam=\ex{1}$), the drop in wave energy is much more dramatic in 2D than in 1D, by roughly one order of magnitude. At cold temperatures ($\delgam\lesssim \ex{5.5}$), the precursor waves are stronger in 1D than in 2D. We attribute this difference in efficiency  at low temperatures to the longitudinal heating induced in 2D by the filamentation mode, which is absent in 1D (see also \app{temp}, for the effect of longitudinal dispersion on the precursor strength). Note that the 1D simulations presented by \citet{plotnikov2019} employed $\delgam=\ex{4}$, for which 1D and 2D results only differ by $\sim 50\%$.    

As regard to the dependence on magnetization, the top panel of \fig{xieff} shows that the normalized wave energy is roughly the same between $\sigma=1$ and $\sigma=3$: for cold and moderate temperatures ($\delgam\lesssim \ex{1.5}$), $\xi_B$ is a few percent, while it drops by nearly two orders of magnitude for $\delgam\gtrsim \ex{1}$. In the range of temperatures where the precursor is efficient, the minimum of the normalized wave energy is attained for $\delgam\sim \ex{2.5}$ in $\sigma=1$ and for $\delgam\sim \ex{3.5}$ in $\sigma=3$. We shall use these  values to roughly distinguish between cold cases and warm cases, which, as shown in \sect{spec} for $\sigma=1$, display different spectral properties. 

From the normalized wave energy in the top panel of \fig{xieff}, we can extract the so-called strength parameter
\be
a=\frac{e\,\delta E_y}{m c\omega}
\ee
At $\sigma\gtrsim 1$, where the typical wave frequency in cold plasmas is $\omega\sim 3\,\sigma^{1/2}\omp$ \citep{plotnikov2019}, we find that $a\sim 0.3\, (\xi_B/\ex{2})^{1/2}(\gamma_0/10)$. The strength parameter could also be measured directly from the transverse motion (along $y$) of the upstream particles  in the field of the wave, since the oscillations in $\gamma\beta_y$ are directly related to the wave strength parameter \citep[e.g.,][]{lyubarsky_06,iwamoto_17}. Notice that we expect the particle transverse oscillations to become relativistic at $\gamma_0\gtrsim 30$. However, we have performed 2D simulations with $\sigma=1$ and $\delgam=\ex{4}$ for pre-shock Lorentz factors up to $\gamma_0=80$ and we do not find that this changes the precursor dynamics or efficiency, in agreement with our earlier 1D results \citep[see figure 8 in][]{margalit_20}. 

The second panel in \fig{xieff} shows the shock velocity in units of the speed of light, as measured in the downstream frame. Its dependence on dimensionality and temperature (in the regime $\delgam\lesssim \ex{1}$ we have explored)  is minimal. The marginal reduction at large $\delgam$ is expected based on MHD jump conditions. The fact that for $\sigma=1$, 1D shocks in very cold plasmas are slightly slower than 2D shocks mirrors the different efficiency in precursor emission: in 1D the precursor takes away a larger fraction of the plasma energy, thereby slowing down the shock.
The dependence on magnetization follows the expectation of MHD jump conditions. Assuming the adiabatic index of a relativistic gas with two degrees of freedom ($\hat{\gamma}_{\rm ad}=3/2$; in fact, our particles only isotropize in the $xy$ plane orthogonal to the magnetic field), the  dimensionless 4-velocity of ultra-relativistic magnetized (i.e., $\gamma_0\gg1$ and $\sigma\gg 1$) shocks is expected to be $\gamma_{\rm sh}\beta_{\rm sh}=(5\sigma/4+7/20)^{1/2}$ \citep[e.g.,][]{petri_lyubarsky_07,plotnikov_18}. At $\sigma\gg1$, this scales as  $\gamma_{\rm sh}\beta_{\rm sh}\propto \sigma^{1/2}$. As we discuss in \sect{spec}, this has important implications for the beaming of the precursor emission.

The data in the first and second panels are used to compute the wave efficiency $f_\xi$ shown in the third panel. This is defined as the fraction of incoming total energy (electromagnetic and kinetic) that is converted into precursor wave energy. In the downstream frame of the simulations, the energy that has flown into the shock per unit area up to time $t$ is
\be
    E_{\rm in}=\gamma_0(1+\sigma)N_0 m c^2 (\beta_0+\beta_{\rm sh}) c t
\ee
where the flux factor $(\beta_0+\beta_{\rm sh})$ accounts for the fact that the shock is moving towards the upstream. The energy converted into precursor waves per unit shock area is 
\be
    E_{\rm out}=\frac{\langle\delta B_z^2 \rangle}{4 \pi} (1-\beta_{\rm sh}) c t
\ee
where we have assumed that the whole region between the shock and the leading edge of the precursor (moving at $c$) is occupied by precursor waves with uniform energy density.
The efficiency is then 
\be
f_{\xi}=\frac{E_{\rm out}}{E_{\rm in}}=\xi_B \left({\sigma \over 1+\sigma} \right) \left( {1-\beta_{\rm sh} \over \beta_0 + \beta_{\rm sh}} \right)
\label{eq:xidef}
\ee
Notice that in the $\sigma\gg1$ limit this scales as $f_{\xi}\propto \xi_B \sigma^{-1}$. In the shock maser scenario for FRBs, this quantifies the fraction of blast wave energy that is converted into precursor wave energy (i.e., the candidate FRB). The ratio of precursor power to blast wave power can be obtained by accounting for the duration of the precursor emission, which in the limit $\sigma\gg1$ is a factor of $\sim \sigma^{-1}$ shorter than the blast wave ejection duration. It follows that in the limit of high magnetizations the ratio of emitted precursor power to blast wave power is $\sim \xi_B$. 

The bottom panel of \fig{xieff} shows that the precursor efficiency in 2D simulations scales  as $f_\xi\sim \ex{3}\sigma^{-1}$ as long as $\delgam\lesssim \ex{1.5}$, whereas it drops abruptly to $f_\xi\lesssim \ex{5}$ for $\delgam\gtrsim \ex{1}$.

%The solid points in the bottom panel indicate the expected value of the momentum variance in a plasma with no precursor. The average value of the x-momentum variance displays very little deviation from the expected value for the hot plasmas such as $kT/mc^2 = 0.1$. This indicates that in hotter plasmas the electron oscillations due to the precursor wave are less than the spread due to thermal motion. The colder plasmas display an increasing deviation from the expected value. In colder plasmas the precursor wave leads to oscillations greater than the thermal spread. This is likely due to the fact that the precursor in the colder plasmas has a larger normalized Poynting flux as evidenced by the top panel. 

%%%%%%%%%%%%%%%%%%%%%%%%%%%%
\begin{figure}
\centering
%\begin{minipage}[b]{.45\textwidth}
\includegraphics[width=1.0\columnwidth,angle=0]{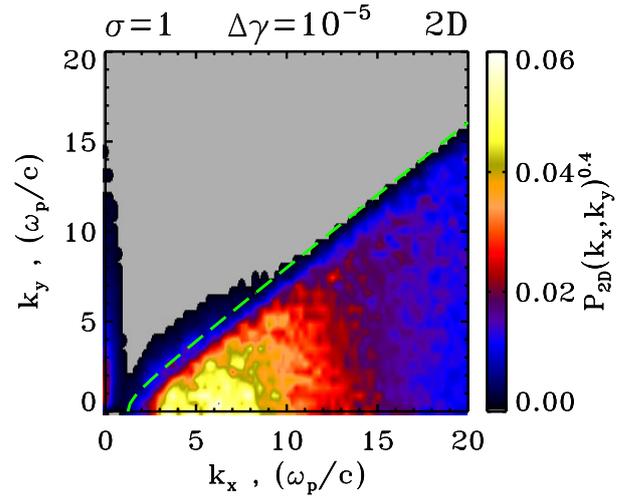}
\caption{2D wavenumber power spectrum $P_{\rm 2D}(k_x,k_y)$ for the 2D simulation with $\sigma=1$ and $\delgam=\ex{5}$, averaged in the time range $2000\leq\omp t\leq2500$. The spectrum  is extracted from the same region ahead of the shock where $\xi_B$ is computed, i.e $5\,\comp<x-\xsh<30\,\comp$, and is normalized such that at each time its integral equals $\xi_B$. The dashed green line indicates the theoretical upper limit in \eq{cutoff} for waves with group speed larger than the shock speed, taking the measured $\beta_{\rm sh}=0.78$.}\label{fig:spec2d5}
\end{figure}

\begin{figure}
%\end{minipage}\qquad
%\begin{minipage}[b]{.45\textwidth}
\includegraphics[width=1.0\columnwidth,angle=0]{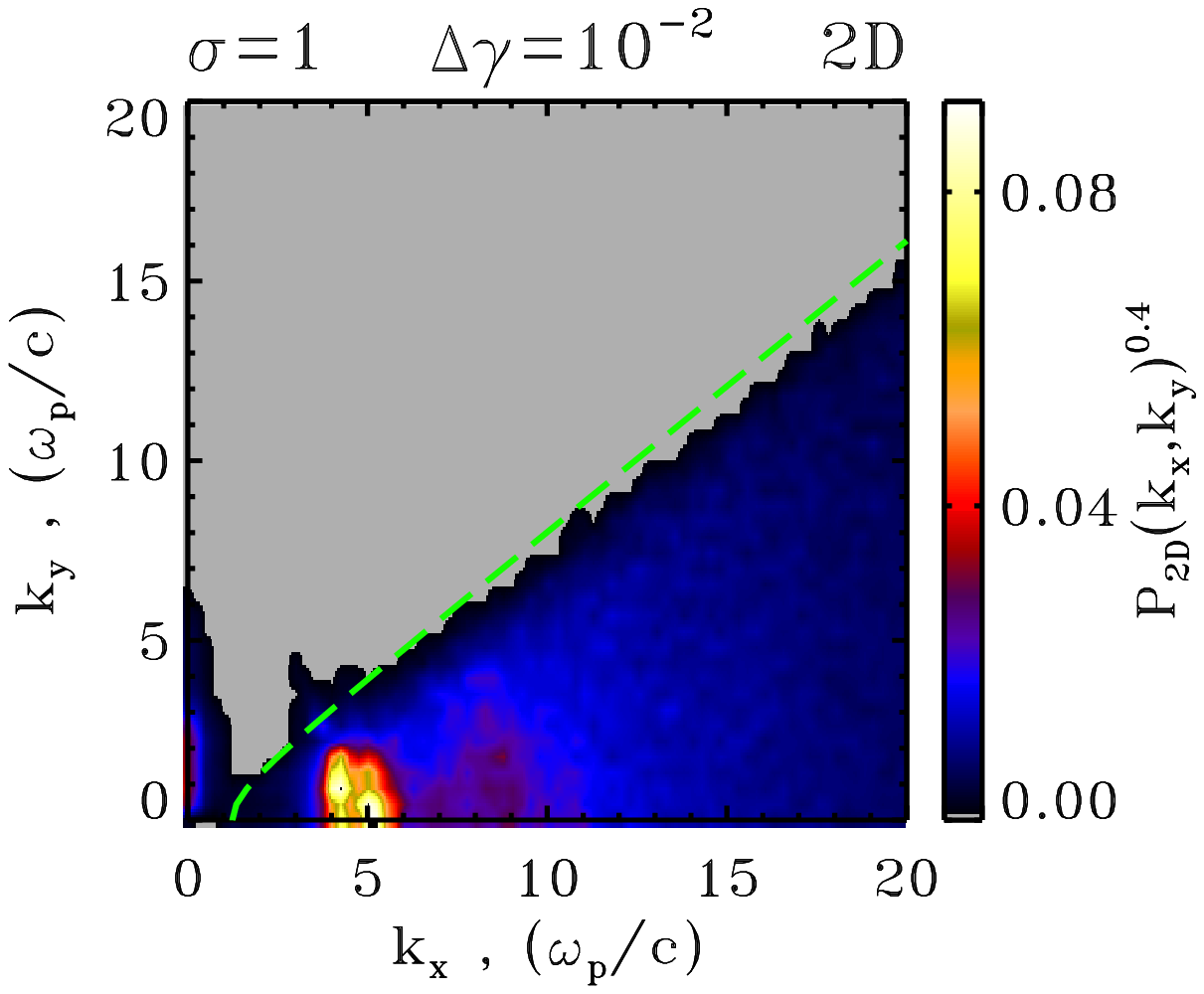}
\caption{As in \fig{spec2d5}, for the 2D simulation with $\sigma=1$ and $\delgam=\ex{2}$, averaged in the time range $2000\leq\omp t\leq2500$. The dashed green line corresponds to \eq{cutoff} taking the measured $\beta_{\rm sh}=0.78$.}\label{fig:spec2d2}
%\end{minipage}
\end{figure}

\begin{figure}
    \centering
    \includegraphics[width=\columnwidth]{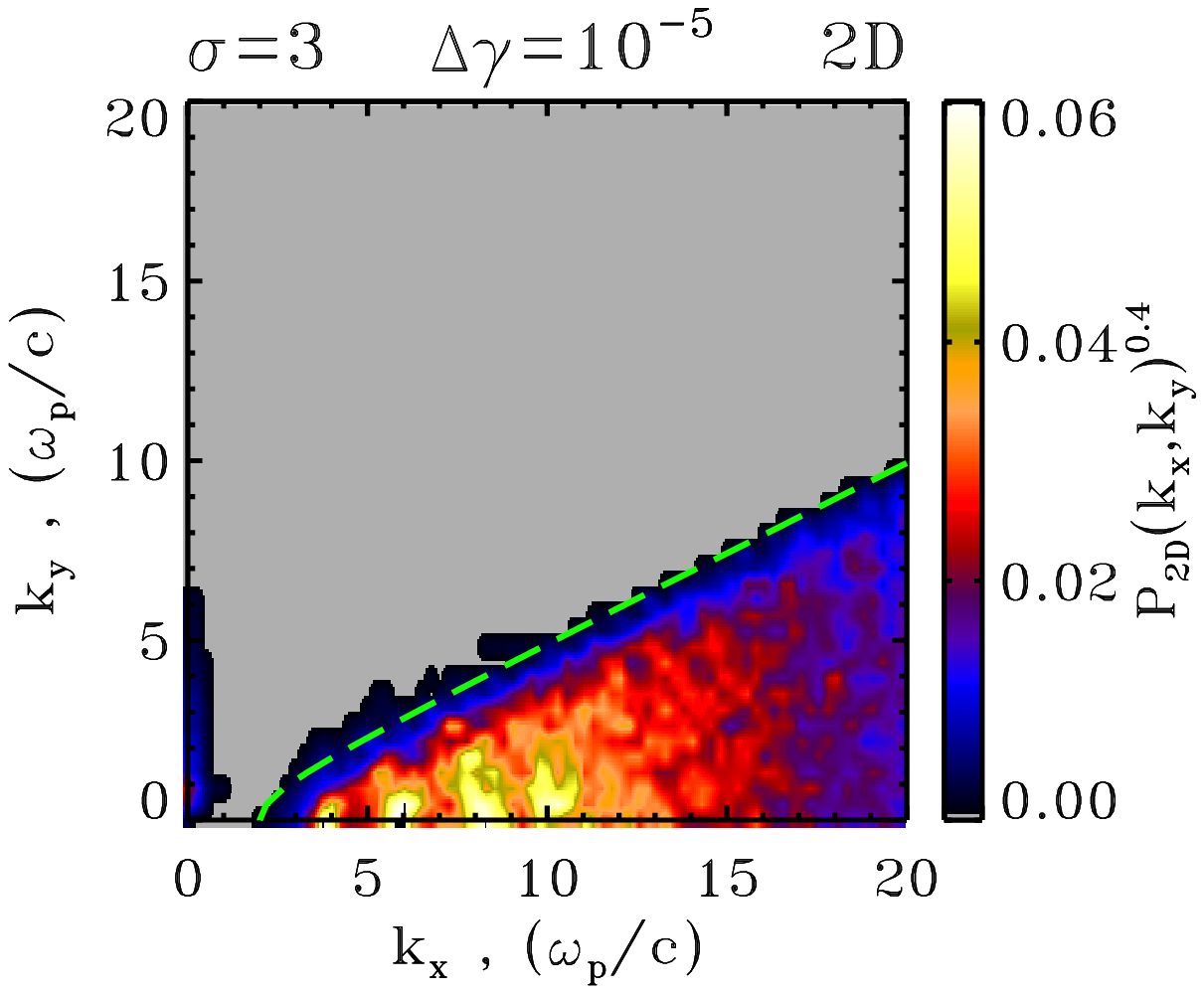}
    \caption{As in \fig{spec2d5}, for the 2D simulation with $\sigma=3$ and $\delgam=\ex{5}$, averaged in the time range $4000\leq\omp t\leq4500$. The dashed green line corresponds to \eq{cutoff} taking the measured $\beta_{\rm sh}=0.89$.}
    \label{fig:spec2d3}
\end{figure}

\begin{figure}
    \centering
    \includegraphics[width=\columnwidth]{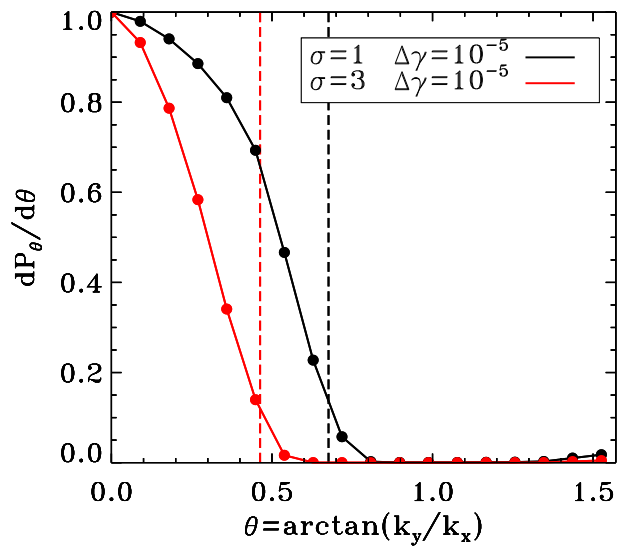}
    \caption{Angular distribution of the precursor power $dP_\theta/d\theta$, where $\theta=\arctan(k_y/k_x)$, obtained from the 2D power spectrum $P_{\rm 2D}(k_x,k_y)$ by integrating along lines of constant $\theta$. We show results for the same simulations shown in Figs.~\fign{spec2d5} and \fign{spec2d3} (see legend): black for $\sigma=1$ and $\delgam=\ex{5}$ (\fig{spec2d5}), red for $\sigma=3$ and $\delgam=\ex{5}$ (\fig{spec2d3}). The curves are normalized to their respective maximum values. The vertical dashed lines (same color coding as the solid lines) represent the boundary $\theta_{\rm crit}=\arctan(1/\gamma_{\rm sh}\beta_{\rm sh})$ defined by \eq{cutoff}, such that only waves with $\theta\leq \theta_{\rm crit}$ can outrun the shock. }
    \label{fig:beaming}
\end{figure}
%%%%%%%%%%%%%%%%%%%%%%%%%%%%

\begin{figure}
\begin{center}
%\vspace{-0.20in}
\includegraphics[width=\columnwidth]{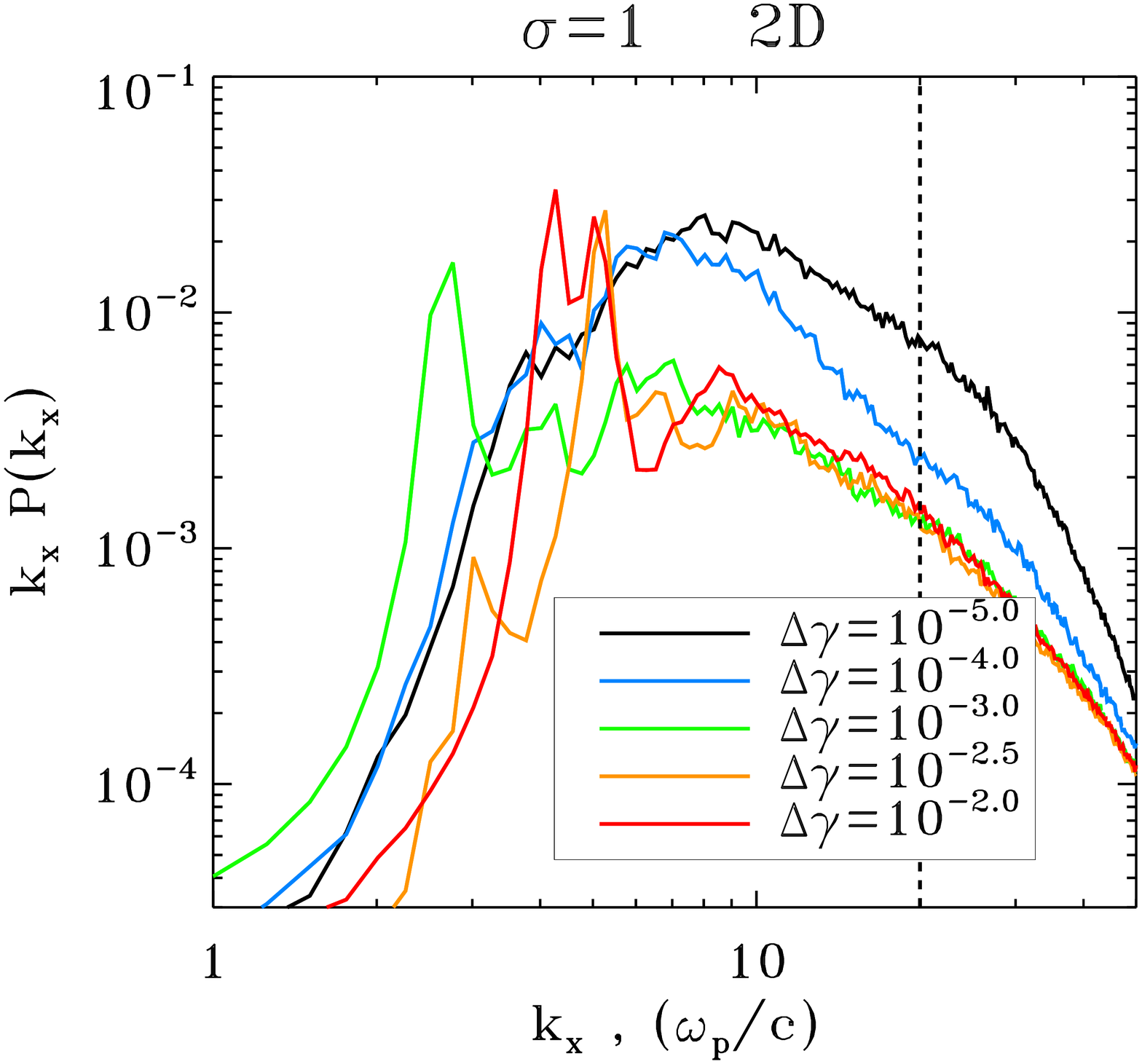}
%\vspace{-0.07in}
\caption{1D wavenumber power spectrum $P(k_x)=\int P_{\rm 2D}(k_x,k_y)\,d k_y$ for 2D simulations with $\sigma=1$ and varying  $\delgam$ (see  legend). The spectrum is extracted from the same region ahead of the shock where $\xi_B$ is computed, i.e., $5\,\comp<x-\xsh<30\,\comp$. At each time, the spectrum is normalized such that its integral equals $\xi_B$. Each curve is obtained by averaging the spectra in the time interval $2000\leq\omp t\leq2500$. The vertical dashed line indicates the boundary $k_{\rm max}=20\,\omp/c$ beyond which our spectra may be affected by numerical artifacts (see \app{num}). The Nyquist wavenumber for our simulations is $k_N\simeq314\,\omp/c$.}
\label{fig:spec1d_comp}
%\vspace{-0.4in}
\end{center}
\end{figure}

%%%%%%%%%%%%%%
%%%%%%%%%%%%%%
%%%%%%%%%%%%%%
\section{Precursor Beaming and Power Spectrum}\label{sect:spec}
To characterize the spectral properties of precursor waves, we have constructed the 2D wavenumber spectrum $P_{\rm 2D}(k_x,k_y)\propto |\delta \tilde{B_z}(k_x,k_y)|^2$, by taking the Fourier transform $\delta\tilde{B_z}(k_x,k_y)$ of the fluctuating magnetic field $\delta{B_z}(x,y)$.  Our spectra are computed in the post-shock frame, by extracting $\delta{B_z}(x,y)$ in the same region ahead of the shock ($5\,\comp< x-\xsh<30\,\comp$) where we compute the precursor efficiency. We will show both the 2D power spectrum $P_{\rm 2D}(k_x,k_y)$ and the $k_y$-integrated 1D power spectrum $P(k_x)=\int P_{\rm 2D}(k_x,k_y)\,d k_y$.\footnote{In \citet{plotnikov2019}, we showed that for 1D simulations, wavenumber spectra ($k_x$-spectra) and frequency spectra ($\omega$-spectra) nearly overlap, when accounting for the dispersion relation of X modes.} The power spectra are normalized such that $\int P_{\rm 2D}(k_x,k_y)\,d k_x dk_y=\int P(k_x)dk_x=\xi_B$.

In Figs.~\fign{spec2d5}-\fign{spec2d3}, we present the 2D power spectrum for three representative simulations, having different values of magnetization and pre-shock temperature. In \fig{spec2d5} we show results for $\sigma=1$ and $\delgam=\ex{5}$ (the cold case described in \fig{shock5} and \fig{pspace5}), in \fig{spec2d2} for $\sigma=1$ and $\delgam=\ex{2}$ (the warm case in \fig{shock2} and \fig{pspace2}), and in \fig{spec2d3} for $\sigma=3$ and $\delgam=\ex{5}$ (a cold case with higher magnetization).

In each of the plots, the power at $k_x\sim 0$ and $k_y\gtrsim \ompc$ is attributed to wave filamentation associated to the density filaments observed in panel (c) of Figs.~\fign{shock5} and \fign{shock2}. Most of the spectral power, however, resides at higher $k_x$, within the region delimited by the green dashed line (defined below). By comparing the figures, one sees that for cold plasmas (\fig{spec2d5} and \fig{spec2d3}) the power is distributed over a wide range of longitudinal wavenumbers ($k_x\sim 5-10\ompc$), whereas for warm plasmas (\fig{spec2d2}) the spectrum is sharply peaked at $k_x\sim 5\ompc$. This will be further discussed below, where we show the $k_y$-integrated spectrum $P(k_x)$.

By comparing \fig{spec2d5} and \fig{spec2d3}, which differ in magnetization, we find that the precursor emission is beamed within a narrower angle $\theta=\arctan(k_y/k_x)$ for larger $\sigma$. As we now discuss, this follows from the requirement that the waves be able to escape ahead of the shock, which moves faster for higher magnetizations (see middle panel of \fig{xieff}).

The dispersion relation of the extraordinary mode (X mode) in cold plasmas in the frame where the background plasma is at rest reads \citep[see, e.g.,][]{hoshino_91}
\begin{eqnarray}
{k^{\prime \prime 2} c^2 \over \omega^{\prime \prime 2}} = 1 - {\omega_{\rm p} ^{\prime \prime 2} \over \omega^{\prime \prime 2} -\sigma \omega_{\rm p} ^{\prime \prime 2}} \, ,
\end{eqnarray}
where double primed quantities are measured in the upstream rest frame. In the limit $\gamma_0^2 \gg \sigma$, the dispersion relation in the downstream frame becomes
\begin{equation}
k^2 c^2 \simeq \omega^2-\omega_{\rm p}^2 \, .
\label{eq:dispersion_DRF}
\end{equation}
%Here, all quantities are expressed in the DRF that moves with $\gamma_0$ Lorentz factor with respect to the URF. 
which is identical to the dispersion relation of a simple electromagnetic wave propagating in an unmagnetized plasma.

The motion of the shock front imposes a cutoff below which the wave cannot escape into the upstream medium. This cutoff is obtained by equating the projection of the group velocity $v_g={\rm d} \omega / {\rm d} k$ onto the shock normal with the shock speed, i.e., $v_g\cos\theta\geq\beta_{\rm sh} c$, which leads to 
\begin{equation}
k_x \geq\gamma_{\rm sh} \beta_{\rm sh}\sqrt{k_y^2 + \frac{\omega_{\rm p}^2}{c^2}}
\label{eq:cutoff}
\end{equation}
This inequality identifies the values of $(k_x,k_y)$ for which the wave can successfully outrun the shock \citep{iwamoto_17}.

This has two main consequences. First, the range of $k_x$ allowed for wave propagation has an absolute lower limit, $k_{x,\rm cut}=\gamma_{\rm sh}\beta_{\rm sh}\ompc$. In the limit of high magnetizations, this scales as $k_{x,\rm cut}\simeq\sigma^{1/2}\ompc$, in agreement with a comparison of \fig{spec2d5} and \fig{spec2d3}. Second, for each given $k_x$, the allowed range of $k_y$ is constrained by \eq{cutoff}, which is indicated by the dashed green lines in Figs.~\fign{spec2d5}-\fign{spec2d3}. At $k_x,k_y\gg\ompc$, this corresponds to precursor emission being confined within an angle $\theta_{\rm crit}= \arctan(1/\gamma_{\rm sh}\beta_{\rm sh})$ from the shock normal. For $\sigma\gg1$, this scales as $\theta_{\rm crit}\simeq \sigma^{-1/2}$. So, precursor waves from more strongly magnetized shocks will be directed closer to the shock normal.

This is demonstrated in \fig{beaming}, by computing the angular distribution of precursor power $dP_\theta/d\theta$, obtained by integrating the 2D wavenumber spectrum $P_{\rm 2D}$ along lines of constant $\theta$. The two curves correspond to the two cases in Figs.~\fign{spec2d5} and \fign{spec2d3} (see legend), and are normalized to their respective peak values. The vertical dashed lines (same color coding as the solid lines) correspond to $\theta_{\rm crit}= \arctan(1/\gamma_{\rm sh}\beta_{\rm sh})$. The plot shows that the precursor emission at $\theta>\theta_{\rm crit}$ is indeed negligible (the power at $\theta=\pi/2$ is contributed by the non-propagating filamentation mode, as described above). As expected, precursor waves in flows with higher magnetizations are more strongly beamed, since $\theta_{\rm crit}\simeq \sigma^{-1/2}$. Moreover, the precursor is even more beamed than the constraint in \eq{cutoff} would prescribe: the width at half maximum of $dP_\theta/d\theta$ is $\sim 0.7\,\theta_{\rm crit}$.

Further comparison of wavenumber spectra among cases with fixed $\sigma=1$ and varying temperatures is presented in \fig{spec1d_comp}, where we show the $k_y$-integrated wavenumber spectrum $P(k_x)$ (more precisely, we show $k_xP(k_x)$ to emphasize where most of the power resides). We focus our discussion on the range $k_x< 20\ompc$ which is robust against variations of numerical parameters, see \app{num}.

Once the precursor efficiency settles to a steady state (see \fig{time}), the spectral shape is also nearly time independent. Regardless of the pre-shock temperature, the spectra share some common features: ({\it i}) the range of longitudinal wavenumbers has a sharp cutoff at $k_{x,\rm cut}\simeq 2\ompc$, which descends from the constraint in \eq{cutoff} for $k_y=0$; ({\it ii}) the spectral shape at high wavenumbers ($k_x\gtrsim 5\ompc$) resembles a power law $P(k_x)\propto k_x^{-2}$.

Despite these similarities, sharp differences exist between cases with different pre-shock temperatures. For cold temperatures, the spectrum peaks at $k_x\sim 5-10\,\omp/c$ and is relatively broad, with fractional width $\sim 1-3$. This is common to all cases with $\delgam\lesssim \ex{3.5}$ (we show $\delgam= \ex{5}$ in black  and $\delgam= \ex{4}$ in blue). In this temperature range, the spectral power at high wavenumbers ($k_x\gtrsim 10\ompc$) gets increasingly suppressed for larger thermal spreads, as discussed by \citet{amato_arons_06}.
For warm temperatures ($\ex{3}\lesssim\delgam\lesssim \ex{1.5}$), the spectrum shows pronounced line-like features with fractional width $\sim 0.2$. The line-like features are located at the low-wavenumber end of the spectrum, at $k_x\sim 3-5\,\omp/c$.  We show spectra for $\delgam= \ex{3}$ (green), $\delgam= \ex{2.5}$ (yellow) and $\delgam= \ex{2}$ (red). The spectrum for $\delgam= \ex{1.5}$ (not shown) is very similar to the $\delgam=\ex{2}$ case, whereas we remind that the precursor efficiency is strongly suppressed  for even hotter temperatures ($\delgam\gtrsim \ex{1}$).

%%%%%%%%%%%%%%%%%%
%%%%%%%%%%%%%%%%%%
%%%%%%%%%%%%%%%%%%
\section{Summary and Discussion}\label{sect:disc}
In this work we have investigated by means of 2D PIC simulations the physics of the precursor waves emitted by perpendicular relativistic electron-positron shocks with out-of-plane upstream fields. We have focused on the high magnetization regime $\sigma\gtrsim 1$ appropriate for magnetar winds, motivated by the shock-powered synchrotron maser scenario proposed for FRBs \citep{lyubarsky_14,murase_16,belo_17,waxman_17,lorenzometzger,belo_19,margalit_20,margalit_20b}. We have explored the efficiency and spectrum of the precursor waves as a function of the pre-shock thermal spread $\delgam=kT/mc^2$ in the range $\delgam=\ex{5}-\ex{1}$. All our simulations have been run for a sufficiently long time ($\gtrsim 4000\, \omp^{-1}$) that the precursor emission has achieved a steady state. Our main results are:
\begin{enumerate}
\item By measuring the fraction $f_\xi$ of total (i.e., electromagnetic and kinetic) incoming energy  that is converted into precursor waves, as computed in the post-shock frame, we can quantify the efficiency of precursor emission. At fixed temperature, the scaling with magnetization $f_\xi\sim 10^{-3}\,\sigma^{-1}$  at $\sigma\gtrsim 1$ is consistent with our earlier 1D results \citep{plotnikov2019}. 
\item At fixed magnetization, the precursor efficiency is nearly independent of temperature as long as $\delgam\lesssim \ex{1.5}$ (with only a modest decrease of a factor of three from $\delgam=\ex{5}$ to $\delgam=\ex{1.5}$), but between $\delgam=\ex{1.5}$ and $\delgam=\ex{1}$ it drops by nearly two orders of magnitude. We have confirmed this result with dedicated 3D simulations (Sironi et al, in prep.) So, shocks propagating in hot plasmas with $\delgam\gtrsim\ex{1}$ are unlikely to power FRBs.
\item For $\sigma\gtrsim 1$, the precursor waves are beamed within a cone of half-opening angle $\theta_{\rm crit}=\arctan(1/\gamma_{\rm sh}\beta_{\rm sh})\simeq \sigma^{-1/2}$ around the shock normal (as measured in the post-shock frame). This stems from the fact that only the waves whose group velocity projected along the shock normal is larger than the shock speed can outrun the shock. More precisely, the width at half maximum of the angular distribution of precursor power is $\sim 0.7\,\theta_{\rm crit}$.
\item For $\sigma=1$, we have compared the power spectrum $P(k_x)$ of precursor waves (integrated over the transverse wavenumber $k_y$) among different $\delgam$.  For cold temperatures, the spectrum peaks at $k_x\sim 5-10\,\omp/c$ and is relatively broad, with fractional width $\sim 1-3$. In contrast, for warm temperatures ($\ex{3}\lesssim\delgam\lesssim \ex{1.5}$) it shows pronounced line-like features with fractional width $\sim 0.2$. The line-like features are located at the low-wavenumber end of the spectrum, at $k_x\sim 3-5\,\omp/c$. For both cold and warm flows, the high-wavenumber part at $k_x>10\,\omp/c$ can be roughly modeled as a power law $P(k_x)\propto k_x^{-2}$.
\end{enumerate}
Our simulations employ 2D computational domains initialized with out-of-plane magnetic fields. This configuration only allows for the excitation of the X mode (and not of the O mode), so in our case the precursor waves are $100\%$ linearly polarized,  with fluctuating magnetic field along the same direction as the upstream mean field. 
%In Appendix B of \citet{plotnikov2019}, we discussed the relative importance of O modes and X modes in 3D shock simulations of  cold plasmas ($\delgam=\ex{4}$), and described that O modes are generally suppressed at $\sigma\gtrsim 1$. 
A  3D investigation of the dependence of the precursor emission (in terms of both O and X modes) on the pre-shock temperature will be presented elsewhere. 
%Our preliminary results confirm that one of the main conclusions of this work --- the fact that the precursor efficiency drops abruptly for hot upstream flows --- also holds in 3D simulations.

The shocks investigated in this work propagate in an electron-positron plasma. The physics of electron-proton shocks will qualitatively differ, as discussed analytically by \citet{lyubarsky_06} and investigated with PIC simulations by  \citet{hoshino_08,sironi_spitkovsky_11a,iwamoto_19}. In magnetized electron-proton flows, the electron synchrotron maser instability generates a train of electromagnetic precursor waves propagating into the upstream. In response to the waves, the guiding-center velocity of the incoming electrons decreases, since they experience relativistic transverse oscillations in the strong field of the wave. Due to their high mass, protons are less affected by the waves, instead proceeding close to their initial velocity. The resulting difference in bulk velocity between the two species generates a longitudinal electric ``wakefield,'' which boosts electrons towards the shock. At relatively early times, this results in a substantial increase in the efficiency of precursor emission \citep{iwamoto_19}. However, the nonlinear evolution of the wakefield inevitably leads to appreciable heating of the upstream electrons, which is likely to deteriorate the precursor efficiency in the long term. Future work will help assess the steady-state efficiency of precursor waves from relativistic electron-proton shocks.

Our work can help unveil the importance of the synchrotron maser as a source of coherent emission in astrophysical plasmas. In particular, our results have implications for models of FRBs that rely on maser emission at ultra-relativistic shocks generated by magnetized pulses launched by an active magnetar \citep{lyubarsky_14,murase_16,belo_17,waxman_17,lorenzometzger,belo_19,margalit_20,margalit_20b}. Most directly, our study is relevant for scenarios where the shock propagates in a strongly magnetized electron-positron wind \citep[e.g.,][]{belo_19}. 

Our findings imply that efficient synchrotron maser emission from electron-positron shocks requires the upstream electrons  to have non-relativistic temperatures ($kT/mc^2\lesssim 0.03$). This poses a direct constraint on both the properties of the pristine medium around the magnetar, as well as on the repetition rate of subsequent FRB-generating shocks, since even an initially cold upstream will be
heated by the passage of the shock, raising
the temperature of the medium into which the next pulse
would collide \citep{lorenzometzger,belo_19,margalit_20}. This implies that FRBs from two successive shocks should be separated by a minimum lag time, as required for the plasma shocked by the first one to adiabatically cool before the arrival of the second shock. If this were not to be the case, the second shock would not be able to produce strong precursor waves (unless the second shock outruns the plasma shocked by the first one). For the first repeater FRB 121102 the minimum observed wait time seems to peak around $100\,{\rm s}$, which in turn can be used to place limits on the shock radius \citep{lorenzometzger}.

Even in the case of a cold upstream plasma, some energy may be transferred to the pre-shock flow by either the propagating precursor waves, or by  incoherent synchrotron emission of shock-heated electrons \citep{lorenzometzger,belo_19}. As regard to the former effect, strictly speaking it should not be genuinely considered  ``heating,'' since at any given point ahead of the shock
the particle distribution has negligible dispersion. However, when integrating over a few wavelengths, the particles will display some ``effective'' longitudinal and transverse momentum spread.\footnote{Here, ``longitudinal'' is along the shock direction of propagation ($x$ direction), while ``transverse'' is along the wave electric field  ($y$ direction).} Since the wave-induced motions are mostly transverse to the direction of wave propagation, the transverse dispersion is significantly larger than the longitudinal one \citep[figure 8 in][]{margalit_20}. In \app{temp}, we show that it is the longitudinal dispersion that primarily controls the maser efficiency. For the FRB conditions envisioned by \citet{margalit_20}, the self-heating of upstream electrons by 
the precursor waves is not expected to suppress the maser efficiency.

Concerning the Compton heating of
upstream electrons by the $\gamma$-rays/X-rays generated via incoherent synchrotron emission at the shock, the requirement that this would not raise the pre-shock temperature above $0.03\, m c^2/k$ can be used, if $\gamma$-rays/X-ray emission is  observed in coincidence with the FRB (as for the recent  Galactic FRB, \citealt{mereghetti_20a,mereghetti_20b}), to place important constraints on the shock radius (see eq.~66 in \citealt{lorenzometzger}).  

We have also shown that for large magnetizations, the precursor waves are beamed within a narrow cone around the shock normal, with half-opening angle $\simeq0.7\, \sigma^{-1/2}$. The limited range of emission
angles implies a reduced range of Doppler factors for the
frequency transformation from the post-shock frame to the observer frame \citep{belo_19}, so  the Doppler transformation of the precursor spectrum to the
observer frame will not smear out narrow features. Our results demonstrate that for warm temperatures ($\ex{3}\lesssim \delgam\lesssim \ex{1.5}$), the low-frequency end of the spectrum shows pronounced line-like features with fractional width $\sim 0.2$. Given that such narrow features will not be smeared out in high-$\sigma$ shocks, our spectra may be consistent with the observed complex, and sometimes narrow-band spectral energy distributions of observed FRBs \citep[e.g.][]{ravi_16,law_17,macquart_19}. 

In addition, in the decelerating shock scenario, the sub-pulses observed within a given burst \citep{hessels_19,andersen_19} may be interpreted as due to individual peaks in the maser spectrum, as they
drift downwards across the observing band \citep{lorenzometzger}. Our spectra for warm flows provide  some ground for this interpretation. However, we caution that in the recently detected Galactic FRB \citep{mereghetti_20a,mereghetti_20b,ridnaia_20}, there are two distinct X-ray peaks
accompanying the two radio peaks, which rather suggests that the two radio peaks are associated with two separate shocks, with each X-ray burst coming from incoherent synchrotron emission of the corresponding shock-heated electrons (\citealt{lorenzometzger,belo_19,margalit_20b}; but see e.g., \citealt{lu_20}, for caveats). In this regard, the results presented in this work require the second shock to outrun the plasma shocked by
the first one, otherwise the shock generating the first peak would heat
the gas too much to get efficient maser emission from the second shock.

\section*{Acknowledgements}
We are grateful to B. Margalit, B. Metzger, J. N\"attil\"a, I. Plotnikov, E. Sobacchi, N. Sridhar and A. Tran for inspiring discussions and  comments on the manuscript. 
This work is supported by NASA ATP 80NSSC18K1104. The  simulations  have  been  performed  at Columbia (Habanero and Terremoto), and with NERSC (Cori) and NASA (Pleiades) resources.

%%%%%%%%%%%%%%%%%%%%%
\appendix

%%%%%%%%%%%%%
\section{Dependence on Numerical Parameters}\label{app:num}
\begin{figure}
    \centering
    \includegraphics[width=\columnwidth]{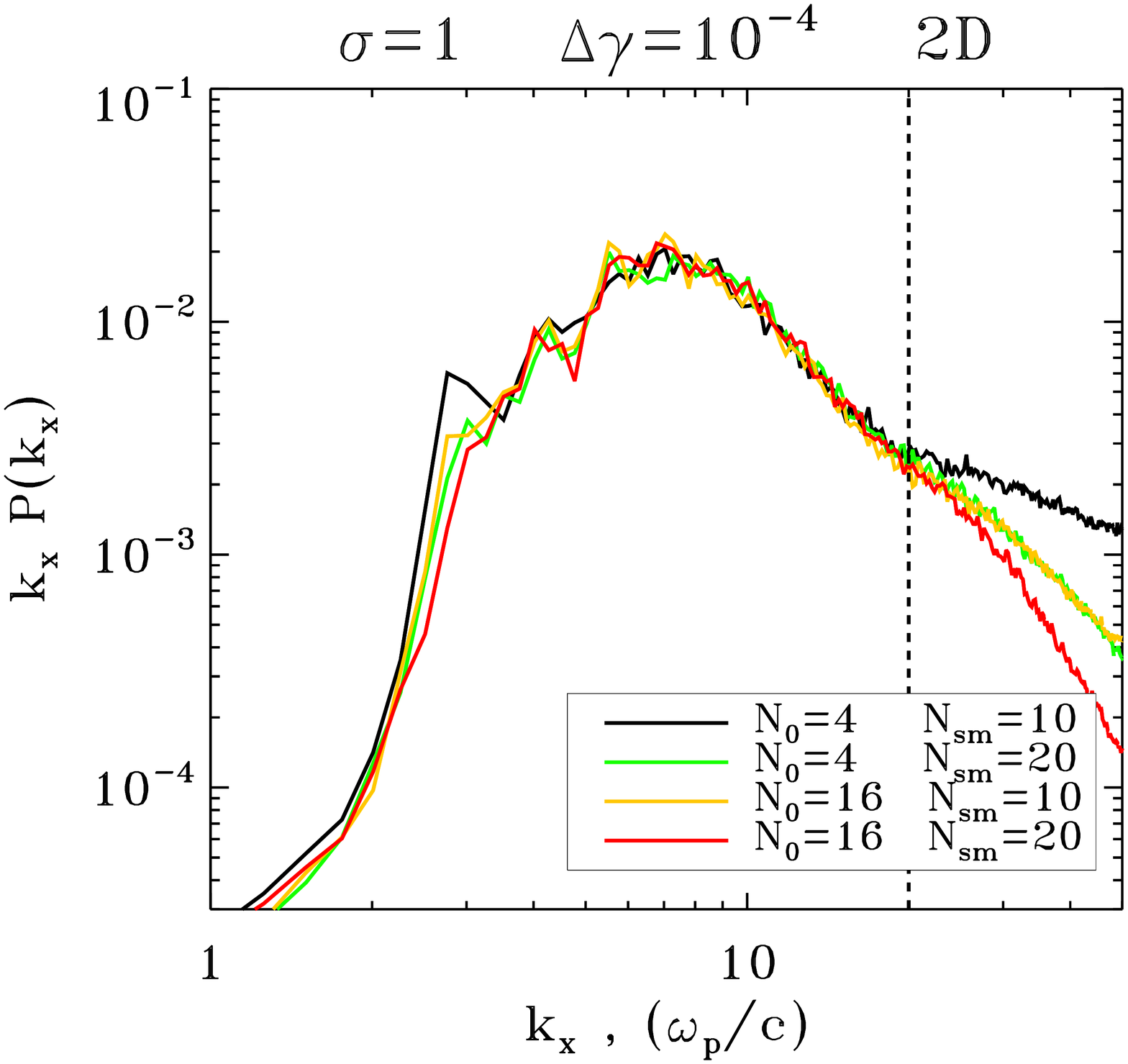}
    \caption{1D wavenumber power spectrum $P(k_x)=\int P_{\rm 2D}(k_x,k_y)\,d k_y$ for 2D simulations with $\sigma=1$ and different choices of particles per cell $N_{\rm 0}$ and number of passes of the smoothing filter of electric currents $N_{\rm sm}$ (see legend). The spectrum is extracted from the same region ahead of the shock where $\xi_B$ is computed, i.e $5\,\comp<x-\xsh<30\,\comp$. At each time, the spectrum is normalized such that its integral equals $\xi_B$. Here, each curve is obtained by averaging the spectra in the time interval $2000\leq\omp t\leq2500$. The vertical dashed line indicates the boundary $k_{\rm max}=20\,\omp/c$ beyond which our spectra differ because of numerical effects.}
    \label{fig:spec1d_num}
\end{figure}

In this Appendix, we describe the dependence of the power spectrum of precursor waves on numerical parameters, for a representative set of 2D simulations with $\sigma=1$ and $\delgam=\ex{4}$. In \fig{spec1d_num}, we show how the spectrum changes when varying the number of particles per cell $N_0$ between 4 and 16, and the number of iterations of the low-pass filter for electric currents between $N_{\rm sm}=10$ and $N_{\rm sm}=20$. Our reference simulations presented in the main body of the paper have $N_0=16$ and $N_{\rm sm}=20$ (red curve in \fig{spec1d_num}).

The figure shows that the spectral shape and normalization at $k\lesssim 20\ompc$ (to the left of the vertical dashed line) are robust to our choice of numerical parameters. On the other hand, the highest-wavenumber part at $k\gtrsim 20\ompc$ varies. A larger number of either particles per cell or filter passes reduces the power at high wavenumbers. It appears that a comparable level of high-wavenumber suppression is achieved with the combinations ($N_{0}=4$, $N_{\rm sm}=20$) and ($N_{0}=16$, $N_{\rm sm}=10$), compare green and yellow lines. In the main body of the paper, we have used the threshold at $k_{\rm max}=20\ompc$ indicated by the vertical dashed line as the upper limit of the wavenumber range where our spectra are robust against numerical artifacts.

%%%%%
\section{Effect of Longitudinal and Transverse Dispersion}\label{app:temp}
In this Appendix, we present dedicated 1D simulations to clarify whether it is the longitudinal or the transverse dispersion that is most detrimental for the efficiency of the precursor emission. We employ $N_0=400$ particles per cell, a spatial resolution of $\comp=112$ cells and a numerical speed of light of $0.5$ cells/timestep. The magnetic field initialization is the same as in the 2D simulations presented in the main body of the paper, and we focus on $\sigma=1$.

\begin{figure}
\includegraphics[width=\columnwidth]{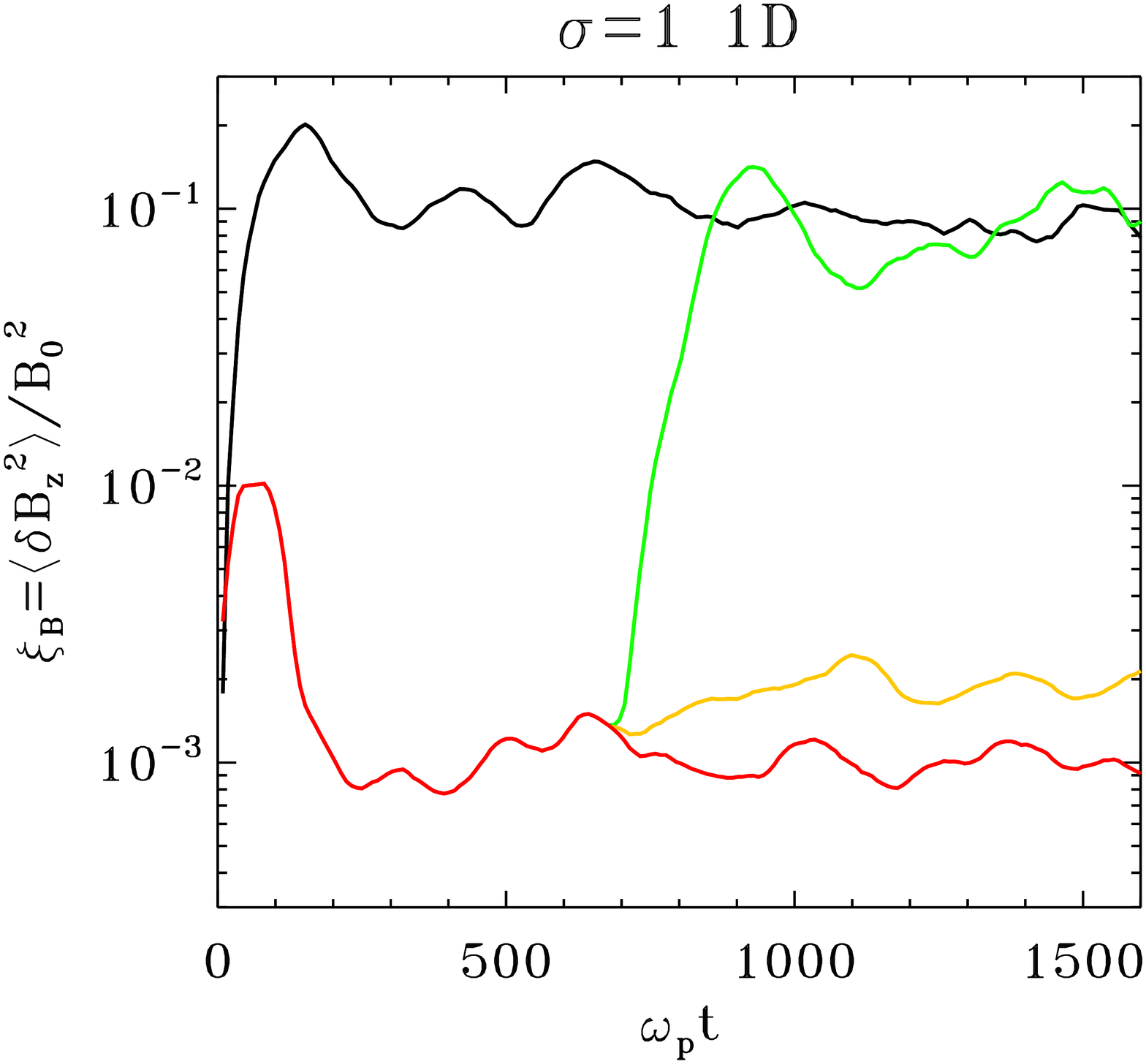}
 \caption{Time evolution of the normalized precursor wave energy, $\xi_B= {\langle \delta B_z^2 \rangle / B_0^2} $ for 1D simulations with $\sigma=1$. The black and red lines respectively correspond to a cold case ($\delgam=\ex{6}$) and a hot case ($\delgam=\ex{0.5}$). The yellow and green lines respectively correspond to cases where we start with a hot thermal spread ($\delgam=\ex{0.5}$), but for $\omp t\geq 700$ we suppress by hand either the longitudinal momentum dispersion (i.e., along $x$) or the transverse one (i.e., along $y$) just ahead of the shock. Here, ``longitudinal'' and ``transverse'' refer to the shock direction of propagation.}
     \label{fig:tempkill}
\end{figure}

\fig{tempkill} shows that the efficiency drops by nearly two orders of magnitude from $\delgam=\ex{6}$ (black) to $\delgam=\ex{0.5}$ (red). Indeed, as \fig{xieff} shows, the drop occurs between $\delgam=\ex{1.5}$ and $\delgam=\ex{1}$. We perform two additional simulations, in which we start with a hot upstream flow ($\delgam=\ex{0.5}$, as for the red line), but for $\omp t\geq 700$ we artificially change its properties right ahead of the shock, in the region $ 4\,\comp \leq x-x_{\rm sh} \leq  22\,\comp$. For the yellow line, we suppress the transverse momentum dispersion (i.e., along $y$), while for the green line we suppress the longitudinal momentum dispersion (i.e., along $x$). If the region where we enforce the suppression were to be far ahead of the shock, gyration around the upstream magnetic field would interchange the transverse and longitudinal motions on a timescale $(\pi/2) \gamma_0\, \sigma^{-1/2} \omp^{-1}$, or equivalently on a distance $(\pi/2) \gamma_0\,\sigma^{-1/2} \comp$. For $\gamma_0=10$ and $\sigma=1$, the choice of suppressing one momentum component between 4 and 22 skin depths is then sufficient to guarantee that the flow entering the shock preserves our imposed temperature anisotropy. 

\fig{tempkill} shows that, if we suppress the transverse dispersion (yellow line), the shock retains the low efficiency of the hot case (red curve). In contrast, if we suppress the longitudinal dispersion (green line), the efficiency increases and eventually settles to the level corresponding to the cold case (black line). We have also verified that at late times the power spectra corresponding to the black and green lines are similar (and the same applies to the spectral comparison of  yellow and red lines).

This demonstrates that, if the upstream plasma were to be continuously heated in an anisotropic way, it is the longitudinal dispersion that determines the efficiency of the precursor emission. Such anisotropic heating may result from the propagation of the precursor waves themselves, since they preferentially induce particle motions in the transverse direction.

%===========================================================================
\bibliographystyle{mnras}
\bibliography{main}
%===========================================================================

\end{document}